\documentclass[10pt,aps,prd,twocolumn,tightenlines,letterpaper,superscriptaddress,nofootinbib,floatfix]{revtex4-2}
\usepackage{amssymb,latexsym}
\usepackage{amsmath,amsbsy,bbm}
\usepackage{epsfig,bm,color,bbold}
\usepackage{graphicx}
\usepackage{xcolor}
\usepackage{cancel}
\usepackage{enumitem}
\begin{document}

\def\a{{\alpha}}
\def\be{{\beta}}
\def\d{{\delta}}
\def\D{{\Delta}}
\def\P{{\Pi}}
\def\p{{\pi}}
\def\e{{\varepsilon}}
\def\ep{{\epsilon}}
\def\g{{\gamma}}
\def\k{{\kappa}}
\def\l{{\lambda}}
\def\L{{\Lambda}}
\def\m{{\mu}}
\def\n{{\nu}}
\def\o{{\omega}}
\def\O{{\Omega}}
\def\S{{\Sigma}}
\def\s{{\sigma}}
\def\t{{\tau}}
\def\x{{\xi}}
\def\X{{\Xi}}
\def\z{{\zeta}}

\def\ol#1{{\overline{#1}}}
\def\c#1{{\mathcal{#1}}}
\def\b#1{{\bm{#1}}}
\def\eqref#1{{(\ref{#1})}}

\def\ed#1{{\textcolor{magenta}{#1}}}


\author{Prabal~Adhikari}
\email[]{$\texttt{adhika1@stolaf.edu}$; KITP Scholar, Kavli Institute for Theoretical Physics, University of California, Santa Barbara 93106.}
 \affiliation{
Physics Department,
        Faculty of Natural Sciences and Mathematics,
        St.~Olaf College,
        Northfield, MN 55057, USA}
\author{Brian~C.~Tiburzi}
\email[]{$\texttt{btiburzi@ccny.cuny.edu}$}
\affiliation{
Department of Physics,
        The City College of New York,
        New York, NY 10031, USA}
\affiliation{
Graduate School and University Center,
        The City University of New York,
        New York, NY 10016, USA}

\title{
QCD Thermodynamics and Neutral Pion in a Uniform Magnetic Field: \\
Finite Volume Effects
} 

\begin{abstract}
We address finite volume effects of lattice QCD calculations in background magnetic fields. 
Using chiral perturbation theory at next-to-leading order, 
volume effects are calculated for thermodynamic quantities: 
the chiral condensate, pressure anisotropy, and magnetization. 
The neutral pion effective action in a finite volume is additionally derived. 
For these charge neutral observables, 
volume and source averaging are shown to capitalize on magnetic periodicity, 
which is the remnant translational invariance of the finite-volume theory.
For a fixed magnetic field strength, 
certain volume and source averaged quantities are independent of the size of the lattice transverse to the magnetic field. 
Despite this simplifying feature, 
finite volume corrections to the magnetic field dependence of the chiral condensate and neutral pion magnetic polarizability can be non-negligible. 
The pressure anisotropy at fixed magnetic flux, 
moreover, 
appears acutely sensitive to the lattice volume. 
\end{abstract}

\maketitle
\section{Introduction}

The study of QCD in background magnetic fields is motivated in part by their phenomenological relevance. 
Large magnetic fields are observed in astrophysical objects, 
with surface magnetization of magnetars reaching nearly $10^{10}$ T%
~\cite{Kong:2022cbk}
and larger fields conjectured for the interiors of these neutron stars%
~\cite{Harding:2006qn}.
Additionally, non-central heavy ion collisions at RHIC and LHC produce immense currents that can result in magnetic fields as large as
$10^{16}$ T%
~\cite{Kharzeev:2007jp,Skokov:2009qp,Deng:2012pc,Inghirami:2019mkc}. 
Fields of similar magnitude may have been produced during the electroweak phase transition after the Big Bang%
~\cite{Vachaspati:1991nm,Vachaspati:2020blt}
and influenced early cosmology of the universe, 
in particular the phase transition from a quark soup to a confined phase of hadronic matter.

In sufficiently strong magnetic fields, 
one can argue that the underlying QCD dynamics becomes weakly coupled%
~\cite{Kabat:2002er}, 
with the zero temperature phase exhibiting magnetic catalysis and anisotropic confinement%
~\cite{Miransky:2002rp}.
For an overview of many aspects of strongly interacting matter in magnetic fields, 
see 
Refs.~\cite{Kharzeev:2013jha,Miransky:2015ava}; 
while, for a review of the QCD phase diagram in an external magnetic field, 
see
Ref.~\cite{Andersen:2014xxa}.
Fortunately, 
non-perturbative quark and gluon interactions can be studied in background magnetic fields using first-principles lattice QCD calculations.
Indeed, 
there has been a wealth of calculations exploring the chiral condensate, QCD thermodynamics, and the phase diagram in magnetic fields, 
{\it e.g.},
Refs.~\cite{DElia:2010abb,DElia:2011koc,Bali:2012zg,Bali:2012jv,Bali:2013esa,Bruckmann:2013oba,Bonati:2013lca,Bonati:2013vba,Bornyakov:2013eya,Bali:2014kia,Endrodi:2015oba,DElia:2018xwo,Ding:2020inp,Ding:2020hxw,DElia:2021tfb,DElia:2021yvk,Ding:2022tqn}. 
Additionally background fields are also of intrinsic interest to QCD, 
because they provide a probe that allows for the investigation of hadron structure and interactions.   
External magnetic fields have been used in lattice QCD calculations to understand, 
{\it e.g.}, 
the masses and magnetic mixing of mesons%
~\cite{Lee:2005dq,Luschevskaya:2014lga,Luschevskaya:2015cko,Bali:2017ian,Bignell:2019vpy,Bignell:2020dze}. 
Despite significant advances in lattice QCD calculations, 
not all computations have been carried out at the physical quark masses, 
nor extrapolated to the continuum and infinite volume limits. 
In particular, 
the reduction of the light quark mass to its physical value on a lattice of a fixed size will lead to an increase in finite volume effects.

A particularly useful tool to systematically investigate finite volume effects that arise in lattice QCD calculations is chiral perturbation theory%
~\cite{Gasser:1983yg,Gasser:1984gg}. 
This low-energy effective theory of QCD is formulated in terms of the pattern of spontaneous and explicit breaking 
of chiral symmetry, 
and incorporates interactions of the emergent pseudo-Goldstone pions.
As pions are the longest-range modes of QCD, 
their dynamics encode effects of the finite volume through their boundary conditions%
~\cite{Gasser:1987zq}.
Observables can be computed order-by-order in a power-counting scheme that treats 
momentum 
$p$
and the pion mass
$m_\p$
as small compared to the chiral symmetry breaking scale
$4 \p F_\p$. 
Additionally one can include external fields in the power counting, 
enabling the consideration of large magnetic fields 
$B$
satisfying
\begin{equation}
\frac{QB}{m_\p^2} \sim 1, 
\quad \text{but with} \quad
Q B \ll ( 4 \p F_\p)^2
,\end{equation}
where 
$Q > 0$
is the charge of the pion. 
Model-independent calculations of QCD properties in magnetic fields have been carried out using 
chiral perturbation theory in infinite volume%
~\cite{Shushpanov:1997sf,Agasian:1999sx,Agasian:2001ym,Cohen:2007bt,Werbos:2007ym,Tiburzi:2008ma,Andersen:2012dz,Andersen:2012zc,Adhikari:2021bou,Adhikari:2021lbl,Adhikari:2021jff,Adhikari:2022vqs}.

In this work,
we study finite volume effects on lattice QCD observables in a homogeneous, background magnetic field. 
The Landau levels of charged particles in magnetic fields exhibit infinite degeneracy due to translational invariance. 
When the volume is finite, 
however, 
only a remnant of the continuous translational invariance remains. 
We explore the consequences of this magnetic periodicity on QCD observables using chiral perturbation theory, 
in which the dominant corrections arise from finite volume modifications to charged pion loops. 
Even for charge neutral quantities, 
magnetic periodicity produces coordinate dependence, 
\emph{e.g.}
the chiral condensate varies in the plane transverse to the magnetic field, 
and the neutral pion experiences a transverse coordinate-dependent potential. 
Volume averaging 
(for bulk thermodynamic quantities)
and source averaging 
(for the neutral pion two-point function)
are shown to remove certain transverse area effects at a fixed value of the magnetic field. 
Despite this simplification, 
finite volume effects on the magnetic field dependence of observables can be non-negligible. 
We find this to be the case of the chiral condensate and neutral pion magnetic polarizability. 
In the case of the pressure anisotropy at fixed magnetic flux, 
moreover, 
we find a substantial finite volume effect for the lowest flux quanta, 
even at 
$m_\p L = 4$.

The presentation is organized as follows. 
In 
Sec.~\ref{sec:FVGF}, 
we review magnetic periodic boundary conditions and obtain the finite volume Green's function of the charged pion, 
which is required for the subsequent chiral perturbation theory calculations. 
Finite volume effects on QCD thermodynamics in a magnetic field are calculated at next-to-leading order in 
Sec.~\ref{sec:observables}.
We begin with the chiral condensate, 
before taking up the free energy, magnetization, and pressure anisotropy. 
In 
Sec.~\ref{sec:action}, 
we determine the finite volume effective action for the neutral pion. 
Consequences of the coordinate-dependent effective potential are discussed, 
and source averaging is shown to restore momentum conservation between the source and sink. 
Various technical details are collected in the Appendices. 
In App.~\ref{s:A}, 
we provide formulas for image sums in terms of elliptic-theta functions. 
Computation of covariant derivatives of the coincident charged pion propagator is given in 
App.~\ref{s:B}. 
The two-point function of the neutral pion is obtained in
App.~\ref{s:C}. 
Lastly, 
a summary of our key results is given in 
Sec.~\ref{sec:conclusion}.

\section{Finite-Volume Green's Function in a Magnetic Field}
\label{sec:FVGF}

The finite-volume Green's function of the charged pion in a magnetic field is central to the chiral perturbation theory calculations in subsequent sections. 
In this section, 
the magnetic periodic boundary conditions required of the Green's function are first reviewed. 
Our treatment follows the quantum mechanical finite-volume problem detailed in 
Ref.~\cite{Al-Hashimi:2008quu}.  
The charged pion propagator is then explicitly constructed using the method of magnetic periodic 
images~\cite{Tiburzi:2014zva}.
The coincident propagator (namely that from a point back to itself) 
is shown to depend on coordinates transverse to the magnetic field,
but maintains the remnant translational invariance of the finite-volume theory.

\subsection{Boundary Conditions}
\label{s:MPBCs}

We work in Euclidean spacetime characterized by the finite-size parameters
$L_\mu = (\beta, L_1, L_2, L_3)$, 
where 
$L_0 = \beta$
is the inverse temperature, 
and 
$L_j$
are the spatial extents. 
Throughout, 
the spatial volume is denoted by 
$V = L_1 L_2 L_3$. 
A uniform magnetic field 
$B \, \hat{x}_3$
can be obtained with the gauge potential 
$\c A_\mu(x) =  A_\mu(x) + B_\mu$, 
where 
$A_\mu(x) = (0,  - B \, x_2, 0, 0)$
and
$B_\mu = (0, \frac{\theta_1}{L_1}, \frac{\theta_2}{L_2}, 0)$
is a constant potential.%
\footnote{
The Green's function for a charged particle depends on the gauge chosen for 
$A_\m(x)$. 
Observables computed in this work, 
however, 
are all charge neutral. 
Thus, 
gauge dependence only occurs at the intermediate stages of calculations, 
with the final results ultimately gauge invariant. 
For example, 
the one-loop corrections with tadpole topology involve the coincident charged pion propagator. 
Such one-loop corrections are necessarily gauge invariant;
but,
in addition to the magnetic field, 
Wilson lines that wrap the compact dimensions are also gauge invariant quantities. 
} 
On a torus, 
$\c A_\m(x)$ 
is periodic up to a gauge transformation
\begin{equation}
\c A_\mu(x + L_2 \hat{x}_2)
= 
\c A_\mu(x) + \partial_\mu \L_2(x)
,\end{equation}
where
$\L_2(x) = - B L_2 x_1$. 
The constant gauge potential 
$B_\mu$
can only be gauged away in infinite volume. 
Note that we have set 
$B_0 = B_3 = 0$ 
for simplicity.

For a complex scalar field 
$\varphi$
of charge 
$Q$ (which we identify with the charged pion starting in Sec.~\ref{s:IIB}), 
we write 
$\c D_\mu \varphi = \left( \partial_\mu + i Q \c A_\mu \right) \varphi$. 
This derivative will be a gauge covariant derivative provided the field transforms as
$\varphi \to e^{ - i Q \L} \varphi$
under the gauge transformation 
$\c A_\mu \to \c A_\mu + \partial_\mu \L$. 
A periodic 
$\varphi$ 
field coupled to 
$\c A_\mu$
includes extra effects due to the finite volume. 
These are best removed from 
$\c A_\mu$
at the cost of modifying the boundary conditions on the scalar field.
The new field
$\phi(x) \equiv e^{- i Q B_\mu x_\mu} \varphi (x)$
removes 
$B_\mu$
from the action,
but leads to twisted boundary conditions
of the form
$\phi(x + L_\mu \hat{x}_\mu) = e^{ - i Q \theta_\mu} \phi(x)$, 
where Einstein's summation convention is suspended in writing this and any subsequent boundary conditions. 
Addressing the effect of non-periodicity of the gauge potential,
moreover, 
requires a gauge transformation at the boundary; 
after which, 
there is an additional factor
\begin{equation}
\phi(x + L_\mu \hat{x}_\mu) = e^{ - i Q \theta_\mu} e^{- i Q \L_\mu(x)} \phi(x)
,\end{equation}
leading to magnetic periodic boundary conditions with a twist. 
It is best to write the phase acquired by wrapping the torus as a Wilson line
\begin{equation}
W_2(x_1) = e^{ - i Q \theta_2} e^{- i Q \L_2(x)} = e^{ - i Q (\theta_2 - x_1 B L_2)}
.\end{equation}
For convenience, 
we define another Wilson line
\begin{equation}
W_1(x_2) = e^{ - i Q (\theta_1 + x_2 B L_1)}
,\end{equation}
for later use.

Consistency of the boundary conditions requires quantization of the magnetic field%
~\cite{tHooft:1979rtg}.
On the one hand, 
we can write
\begin{eqnarray}
\phi(x + L_1 \hat{x}_1 + L_2 \hat{x}_2)
&=&
e^{ - i Q \theta_1}
\phi(x + L_2 \hat{x}_2)
\notag \\
&=&
e^{ - i Q \theta_1}
W_2 (x_1) \phi(x)
,\end{eqnarray}
while on the other, 
we see
\begin{eqnarray}
\phi(x + L_1 \hat{x}_1 + L_2 \hat{x}_2)
&=&
W_2 ( x_1 + L_1)
\phi(x + L_1 \hat{x}_1)
\notag \\
&=&
e^{ - i Q \theta_1}
W_2 (x_1 + L_1) \phi(x)
.\end{eqnarray}
These are equivalent boundary conditions provided
\begin{equation}
e^{ - i Q L_1 L_2 B} 
= 1
\quad
\longrightarrow
\quad
Q B = \frac{2 \pi N_\Phi}{L_1 L_2}
.\end{equation}
The magnetic field strength is thus determined in terms of the integer flux quantum 
$N_\Phi$. 
The flux quantum determines the size of the remnant translational invariance group, 
which is
$\mathbb{Z}_{N_\Phi}$~\cite{Al-Hashimi:2008quu}.
For 
$n \in \mathbb{Z}_{N_\Phi}$, 
we see this invariance manifested in the translational properties of the Wilson lines
\begin{eqnarray}
W_1 \left(x_2 + \tfrac{n}{N_\Phi} L_2 \right)
&=&
W_1 (x_2),
\notag \\
W_2 \left(x_1 + \tfrac{n}{N_\Phi} L_1 \right)
&=&
W_2 (x_1)
\label{eq:DTI}
.\end{eqnarray}
This translational invariance is 
in addition to periodicity, 
under which we have
$W_1(x_2 + n L_2) = W_1 (x_2)$
and
$W_2(x_1 + n L_1) = W_2(x_1)$,
for all 
$n \in \mathbb{Z}$.

\subsection{Charged Pion Propagator}
\label{s:IIB}

Identifying the complex scalar field with the charged pion 
$\phi(x) = \pi^+(x)$, 
the propagator for the charged scalar takes the form
\begin{eqnarray}
G_+(x',x) \equiv \big\langle \pi^{+}(x') \pi^{-}(x) \, \big\rangle
\label{eq:charged}
,\end{eqnarray}
where the angled brackets denote vacuum expectation values. 
Adopting a quantum mechanical notation%
~\cite{Schwinger:1951nm}, 
we have
\begin{eqnarray}
G_+(x',x)
&=&
\langle x' | \frac{1}{-D_\mu D_\mu + m^2} | x \rangle
,\end{eqnarray}
where the covariant derivative
$D_\mu = \partial_\mu - i Q A_\mu$  
acts on the negatively charged pion,
and 
$m$ 
is the tree-level pion mass. 
As such, 
the propagator obeys the Green's function equation 
\begin{equation}
\left(
-D_\mu D_\mu + m^2
\right)
G_{+}(x',x)
= 
\prod_{\mu = 0}^3
\delta_{L_\mu}(x'_\mu-x_\mu)
\label{eq:GreenEQ}
,\end{equation}
where each Dirac delta-function is that having compact support
$x_\mu \in [\, 0, L_\mu)$, 
for each 
$\m$. 
The propagator can be obtained with the help of the proper-time integral
\begin{eqnarray}
G_+(x',x)
&=&
\int_0^\infty ds \, e^{ - s m^2} 
\notag \\
&& \phantom{sp}
\times G_{\parallel} (x'_\parallel, x_\parallel | s)
\, G_\perp (x'_\perp, x_\perp | s)
,\label{eq:GFn}
\end{eqnarray}
where 
$\perp$
denotes the spatial directions transverse to the magnetic field, 
and
$\parallel$
denotes the Euclidean time direction and the field direction. 
The latter directions give rise to a contribution to the proper-time integral having the form
\begin{eqnarray}
G_{\parallel} (x'_\parallel, x_\parallel | s)
&\equiv&
\langle x'_\parallel | \, e^{ \partial_\parallel^2 s}\,  | x_\parallel \rangle
=
\frac{1}{4 \pi s} \sum_{\nu_\parallel} e^{ - \frac{\Delta x_\parallel^2}{4s}}
,\end{eqnarray}
where
$\Delta x_\mu = x'_\mu + \nu_\mu L_\mu - x_\mu$, 
for each value of  
$\mu$.
The sum over 
$\nu_{\parallel} = ( \nu_0, \nu_3 )$
includes all images in the $x_0$- and $x_3$-directions, 
namely
$\sum_{\nu_\parallel} = \sum_{\n_0 = -\infty}^\infty \sum_{\n_3 = -\infty}^\infty$. 
As such, 
$G_{\parallel} (x'_\parallel, x_\parallel | s)$ 
is periodic in both 
$\parallel$-directions.

To complete the specification of the propagator, 
we need the contribution to the proper-time integral from the transverse directions
\begin{eqnarray}
G_\perp(x'_\perp, x_\perp | s)
&\equiv&
\langle x'_\perp | e^{D_\perp^2 s} | x_\perp \rangle
.\end{eqnarray}
On account of Sec.~\ref{s:MPBCs}, 
this contribution obeys magnetic periodic boundary conditions with a twist: 
\begin{eqnarray}
G_\perp(x'_\perp + L_1 \hat{x}_1 , x_\perp | s ) 
&=&
e^{- i Q \theta_1}
G_\perp(x'_\perp, x_\perp | s ), 
\notag \\
G_\perp(x'_\perp + L_2 \hat{x}_2 , x_\perp | s ) 
&=&
W_2(x'_1)
G_\perp(x'_\perp, x_\perp | s )
\label{eq:BC1}
,\end{eqnarray} 
along with
\begin{eqnarray}
G_\perp(x'_\perp, x_\perp + L_1 \hat{x}_1 | s ) 
&=&
e^{+ i Q \theta_1}
G_\perp(x'_\perp, x_\perp | s ),
\notag \\
G_\perp(x'_\perp, x_\perp + L_2 \hat{x}_2 | s ) 
&=&
W^*_2(x_1)
G_\perp(x'_\perp, x_\perp | s )
\label{eq:BC2}
.\end{eqnarray} 
The transverse contribution to the propagator can be related to its infinite-volume counterpart
$G^\infty_\perp$
through a sum over images
constructed to satisfy the boundary conditions given in 
Eqs.~\eqref{eq:BC1} and \eqref{eq:BC2}. 
Writing 
$\n_\perp = ( \n_1, \n_2 )$
with 
$\sum_{\n_\perp} = \sum_{\n_1 = - \infty}^\infty \sum_{\n_2 = - \infty}^\infty$, 
the transverse contribution to the propagator is
\begin{eqnarray}
G_\perp(x'_\perp, x_\perp | s )
&=&
\sum_{\n_\perp}
e^{ i Q \theta_1 \n_1}
\left[
W_2^*(x'_1)
\right]^{\n_2}
\notag \\
&& \phantom{sp}
\times
G^\infty_\perp(x'_\perp + \n_\perp L_\perp, x_\perp | s )
\label{eq:GFn_perp}
,\end{eqnarray}
where the infinite-volume propagator has the form
\begin{eqnarray}
G^\infty_\perp(x'_\perp, x_\perp | s )
\,{=}\,
\c W(x'_\perp, x_\perp) \,
e^{ - \tfrac{Q B (x'_\perp - x_\perp)^2}{4 \tanh Q B s}}
\tfrac{Q B}{4 \p \sinh Q B s} 
.\notag \\
\end{eqnarray}
The infinite-volume propagator is not translationally invariant due to the factor 
$\c W(x'_\perp, x_\perp)$. 
One way to write this factor is as an inverse Wilson line%
\footnote{
The fact that an inverse Wilson line appears guarantees translational and gauge invariance would be maintained by a modified propagator that is defined to include the corresponding Wilson line between the separated charged pion fields in Eq.~\eqref{eq:charged}. 
Such a modified propagator, 
however, 
depends on the path chosen for the Wilson line, 
with the equivalent paths known for pointlike particles. 
} 
evaluated on the straight-line path from 
$x_\perp$
to 
$x'_\perp$, 
namely
\begin{eqnarray}
\c W(x'_\perp, x_\perp)
&=&
e^{- i Q \int_{x_\perp}^{x'_\perp} dz_\mu \, A_\mu(z)}
\notag \\
&=&
e^{\frac{i Q B}{2} (x'_1 - x_1)(x'_2 + x_2)}
.\end{eqnarray} 
With the charged pion propagator fully specified, 
it is straightforward
to confirm that the solution
Eq.~\eqref{eq:GFn}
indeed satisfies the Green's function relation appearing in
Eq.~\eqref{eq:GreenEQ}%
~\cite{Tiburzi:2014zva}.

\subsection{Coincident Propagator}
\label{s:IIC}

In the calculations that follow
(the chiral condensate, free energy, and derivative-free tadpole contributions to the neutral pion effective action), 
the charged pion propagator from a point back to itself is required. 
Due to the remnant translational invariance, 
such coincident contributions are not coordinate independent. 
Instead, 
they depend on the transverse coordinates, 
but maintain   
$\mathbb{Z}_{N_\Phi}$
translational invariance.

The charged pion propagator from a point 
$x_\mu$
back to itself is straightforward to evaluate
using Eq.~\eqref{eq:GFn}.
Not surprisingly, 
it can be expressed in terms of a sum over images
\begin{eqnarray}
G_+(x, x)
&=&
\sum_{\n_\mu}
f(\n_\perp, x_\perp)
\,
g_+(\n_\m)
\label{eq:coin}
,\end{eqnarray}
where the coordinate dependence appears in the phase function
\begin{eqnarray}
f(\n_\perp, x_\perp)
&=&
(-1)^{N_\Phi \n_1 \n_2 } 
\left[
W_1^*(x_2)
\right]^{\n_1}
\left[
W_2^*(x_1)
\right]^{\n_2}
.\,\,\,\label{eq:fperp}
\end{eqnarray}
Notice that each transverse image appearing in the sum is accompanied by the corresponding Wilson lines through the function 
$f(\nu_\perp,x_\perp)$, 
and maintains  
$\mathbb{Z}_{N_\Phi}$
translational invariance via Eq.~\eqref{eq:DTI}. 
Each image, 
moreover, 
has a sign determined by
$(-1)^{N_\Phi \n_1 \n_2}$, 
which can be attributed to a finite volume Aharonov-Bohm effect%
~\cite{Tiburzi:2014zva}.
The remaining image dependence appears under the coordinate-independent, proper-time integral
\begin{eqnarray}
g_+(\n_\m)
{=}
\int_0^\infty ds 
\frac{e^{-s m^2}}{(4\pi s)^2}
\frac{Q B s}{\sinh Q B s} 
e^{- \frac{(\n_\parallel L_\parallel)^2}{4 s} - \frac{Q B (\n_\perp L_\perp)^2}{4 \tanh Q B s}}
.\notag\\
\label{eq:g+}
\end{eqnarray}

The proper-time integral for the term with
$\n_\m = (0, 0, 0, 0) \equiv 0_\m$
is divergent in the ultraviolet 
$s \ll 1$, 
however, 
this is the only term that survives the infinite-volume limit. 
The ultraviolet behavior of the infinite-volume integrand, 
moreover, 
is independent of the magnetic field. 
To regulate the divergence,
we introduce a proper-time cutoff 
$s_0 \ll 1$
into the offending term
\begin{equation}
g_+(0_\m) \equiv \int_{s_0}^\infty ds \, \frac{e^{-s m^2}}{(4\pi s)^2}
\frac{Q B s}{\sinh Q B s} 
\label{eq:offending}
.\end{equation}
Accordingly, 
the one-loop calculations can be renormalized by a magnetic field independent, infinite-volume subtraction of the form
\begin{equation}
g_+(0_\m) \to g^\infty_+ = \lim_{s_0 \to 0} \left[ g_+(0_\m) - \int_{s_0}^\infty ds \, \frac{e^{-s m^2}}{(4\pi s)^2} \right]
,\end{equation} 
where the subtracted term arises from renormalizing the magnetic field independent parameters of the infinite-volume theory.%
\footnote{
It is instructive to write out the term introduced for the subtraction. 
Removing the multiplicative factor of 
$(4 \p)^{-2}$, 
we have
$$
\int_{s_0}^\infty ds \, \frac{e^{-s m^2}}{s^2}
=
\frac{1}{s_0} 
+
m^2 \left[ \log (s_0 \, m^2) + \g_E - 1
\right]
+
\c O (s_0)
.$$
The divergence is absorbed by a 
$- \frac{1}{s_0}$ 
counterterm that is pion mass independent.
The $s_0$-dependence of the chiral logarithm is compensated by the contribution from a low-energy constant.
Schematically, 
we write such a contribution to the above as 
$m^2 \, \ell(s_0)$.
The low-energy constant then satisfies the renormalization group equation 
$s_0 \frac{d}{ds_0} \ell(s_0) = -1$
to keep the net result 
$s_0$-independent.  
As our concern is with the magnetic field and finite volume dependence of observables, 
we renormalize the chiral corrections into the parameters, so that they take their physical values 
in the combined zero-field and infinite-volume limit. 
}
The behavior with respect to the magnetic field and finite volume is hence ultraviolet finite.

The function 
$g^\infty_+$
describes the magnetic field dependence of the coincident propagator in the infinite-volume limit. 
The proper-time cutoff can be removed 
$s_0 \to 0$, 
leading to 
\begin{equation}
g^\infty_+
=
\int_{0}^\infty ds \, \frac{e^{-s m^2}}{(4\pi s)^2}
\left[
\frac{Q B s}{\sinh Q B s} 
{-}\,1 \right]
\equiv
\frac{Q B}{(4 \p)^2} 
\, \c I
\left( \frac{m^2}{QB} \right)
\label{eq:g+inf}
.\end{equation}
The required Laplace transform 
$\c I (\a)$
is defined to be dimensionless. 
It is related to the Hurwitz zeta-function, 
and can be evaluated in terms of the digamma function%
~\cite{Cohen:2007bt}
\begin{equation}
\c I (\a)
=
\a ( 1 - \log \tfrac{\a}{2}) + 2 \log \Gamma(1+ \a) - \log 2 \p
\label{eq:Ialpha}
.\end{equation}

With the infinite-volume limit removed, 
the coincident propagator in 
Eq.~\eqref{eq:coin}
simply becomes
\begin{eqnarray}
\ol G_+(x, x)
&=&
\sum_{\n_\mu \neq \, 0_\m}
f(\n_\perp, x_\perp)
\,
g_+(\n_\m)
\label{eq:Gbar+}
,\end{eqnarray}
and includes effects of finite volume, 
both with and without the magnetic field. 
In what follows, 
the sum over images is useful for analytic manipulations, 
however, 
a more economical expression is needed for numerical evaluations. 
For this purpose, 
the image sums can be expressed in terms of 
Jacobi elliptic-theta functions, 
resulting in the formula
\begin{eqnarray}
\ol G_+(x,x)
=
\int_0^\infty ds \, \frac{e^{- s m^2}}{(4 \p s)^2}
\frac{QB s}{\sinh QB s}
\Big[
\Theta (x_\perp | s)
{-}\,1
\Big]
.\label{eq:G+fin}
\end{eqnarray}
The expression for 
$\Theta(x_\perp|s)$
is somewhat lengthy and
appears in 
App.~\ref{s:A}. 
In infinite volume, 
we have the limit
$\Theta(x_\perp|s) \to 1$; 
and, 
this finite-volume effect appropriately vanishes.

One-loop corrections in chiral perturbation theory also arise from neutral pion contributions. 
While these contributions do not produce magnetic field dependence at the order we work, 
they lead to volume dependence, 
which we discuss for completeness.
The neutral pion propagator is simply the charged pion propagator evaluated at 
$Q = 0$. 
Using the expression from above, 
we have the coincident, neutral pion propagator 
\begin{equation}
G_0(0,0) 
\equiv 
G_0(x,x)
=
\sum_{\n_\m}
g_0 (\n_\m)
,\end{equation}
where the contribution from a periodic image 
$\n_\m$
is
\begin{eqnarray}
g_0(\n_\m)
=
\int_0^\infty ds \, \frac{e^{-s m^2}}{(4\pi s)^2} \,
e^{- \frac{(\n_\m L_\mu)^2}{4 s}}
.\end{eqnarray}
This result is translationally invariant, 
for which we employ the notation 
$G_0(0,0)$.

Using the infinite-volume renormalization scheme,
one has 
$\ol g_0 (0_\m) = 0$
and the sum over all 
$\n_\m$
is replaced by a sum over 
$\n_\m \neq 0_\m$, 
for which the infinite-volume limit accordingly vanishes. 
The renormalized coincident propagator 
$\ol G_0 (0,0)$
can be economically written in terms of Jacobi elliptic-theta functions 
\begin{eqnarray}
\ol G_0(0,0) 
=
\int_0^\infty ds \, \frac{e^{- s m^2}}{(4 \pi s)^2}
\Big[
\Theta_0(s)
-1
\Big]
\label{eq:G0fin}
,\end{eqnarray} 
where 
$\Theta_0(s)$
is defined in 
Eq.~\eqref{eq:Theta0s}. 
The difference of 
$\ol G_0(0,0)$
compared to 
$G_0(0,0)$
amounts to the subtraction of unity in the proper-time integral. 
This subtraction arises due to the absence of 
$\n_\m = 0_\m$
in the sum, 
and ensures that the proper-time integral converges in the ultraviolet. 
In infinite spacetime volume, 
$\Theta_0(s) = 1$
and 
$\ol G_0 (0,0)$
appropriately vanishes.

\section{Finite-Volume Thermodynamics in a Magnetic Field}
\label{sec:observables}

Having spelled out the charged pion propagator in a magnetic field, 
finite-volume thermodynamic quantities can be computed using chiral perturbation theory. 
Specifically, 
we address the chiral condensate, free energy, magnetization and pressure anisotropy. 
Although we can readily evaluate their temperature dependence for temperatures that are small compared to the chiral symmetry breaking scale, 
we show results at zero temperature ($\be = \infty$) in what follows.

\subsection{Chiral Condensate}
\label{s:cond}

Due to the remnant translational invariance in a magnetic field, 
the chiral condensate in finite volume is a local quantity, 
which we denote by
$\langle \, \ol \psi(x) \psi(x) \, \rangle$. 
It can be obtained from the partition function 
$Z$
by functional differentiation with respect to a local scalar source 
$S(x)$,%
\footnote{
We use an isoscalar source, because there is no isospin breaking in the chiral condensate to the order we are working. 
} 
namely
\begin{align}
\langle \, \ol \psi(x)\psi(x)\, \rangle
=
-
\frac{\delta \log Z}{\delta S(x)}
.\end{align}
When the source is replaced by a uniform value, 
such as the quark mass
$s(x) \to m_q$, 
partial differentiation produces the volume averaged condensate
\begin{equation}
\langle \, \ol \psi \psi \, \rangle
= 
- \frac{1}{\be V} \frac{\partial \log Z}{\partial m_q}
\label{eq:psibarpsi}
,\end{equation} 
which can also be obtained by directly computing the spacetime average of the local condensate 
\begin{equation}
\langle \, \ol \psi \psi \, \rangle = \frac{1}{\be V} \int d^4x \, \langle \, \ol \psi(x)\psi(x)\, \rangle
.\end{equation} 
In scenarios with translational invariance, 
the local and average condensates are equal.
In a magnetic field at finite volume, 
however, 
they are no longer equal as
$\langle \, \ol \psi(x)\psi(x)\, \rangle$
depends on 
$x_\perp$ 
due to the remnant 
$\mathbb{Z}_{N_\Phi}$ 
translational invariance.

In chiral perturbation theory for two flavors%
~\cite{Gasser:1983yg}, 
the leading-order value and one-loop correction to the chiral condensate arise from terms in the Euclidean action density 
\begin{eqnarray}
\c L
&=&
S \, \lambda
\left[ 1 -  \frac{1}{2F^2} (\p^0)^2 - \frac{1}{F^2}\pi^+ \pi^- \right]
,\end{eqnarray}
where $x$-dependence of the scalar source and pion fields is treated as implicit. 
In the action density, 
$F \approx 92 \, \texttt{MeV}$ 
is the chiral-limit value of the pion decay constant, 
and
$\lambda < 0$
is the chiral-limit value of the condensate. 
Taking the action density with 
$S(x) \to m_q$, 
the Gell-Mann--Oakes--Renner relation 
$m^2 F^2 = - m_q \, \lambda$
is found by inspection. 

Using the action density, 
the local chiral condensate can be determined at one-loop order from the neutral and charged pion ring diagrams.
The result has the schematic form
\begin{eqnarray}
\langle \, \ol \psi(x) \psi(x) \, \rangle
=
\lambda \left[ 1 - \frac{G_0(0,0)}{2 F^2}  - \frac{G_+(x,x)}{F^2}  \right]
,\qquad
\end{eqnarray}
which omits the renormalization and contributions from low-energy constants. 
These can be handled with the zero-field, infinite-volume renormalization scheme described in Sec.~\ref{s:IIC}.
Denoting 
$\langle \, \ol \psi \psi \, \rangle_{0}$
as the zero-field, infinite-volume limit of the condensate, 
we thus have
\begin{eqnarray}
\frac{
\langle \, \ol \psi(x) \psi(x) \, \rangle}
{
\langle\, \ol \psi \psi \,\rangle_{0}
}
&=&
1 - \frac{\ol G_0(0,0)}{2 F_\p^2}  - \frac{g_+^\infty + \ol G_+(x,x)}{F_\p^2} 
.\quad
\label{eq:Rperp1loop}
\end{eqnarray}
The chiral-limit value of the pion decay constant 
$F$
has been replaced with its physical value
$F_\p$, 
as the difference is beyond the order we work. 
This is similarly done in the propagators, 
where we replace the leading-order pion mass 
$m$, 
with its physical value 
$m_\p$. 
In the zero-field limit, 
the expression for 
$\langle \, \ol \psi(x) \psi(x) \, \rangle$
in Eq.~\eqref{eq:Rperp1loop}
reproduces the $p$-regime finite volume effect on the condensate%
~\cite{Gasser:1987zq};
whereas,
in the infinite-volume limit, 
it reproduces the leading magnetic field dependence of the condensate%
~\cite{Cohen:2007bt}.

Our primary concern is with the finite volume effect in nonzero magnetic fields. 
To this end, 
we subtract the infinite-volume limit 
$\langle \, \ol \psi \psi \, \rangle^\infty = \langle \, \ol \psi \psi \, \rangle_0 \left(1 - g_+^\infty / F_\p^2\right)$, 
and form the condensate ratio
\begin{equation}
R(x_\perp)
= \frac{\langle \, \ol \psi(x) \psi(x) \, \rangle - \langle \, \ol \psi \psi \, \rangle^\infty}{\langle \, \ol \psi \psi \, \rangle_0}
\label{eq:Ratio}
.\end{equation}
In terms of coincident propagators of neutral and charged pions, 
we have
\begin{eqnarray}
R(x_\perp)
&=&
- \frac{1}{F_\p^2}
\left[
\frac{1}{2} \,
\ol G_0(0,0)
+
\ol G_+(x,x)
\right]
.\quad
\end{eqnarray}
Using 
Eqs.~\eqref{eq:G+fin} and \eqref{eq:G0fin} 
for the coincident propagators, 
the finite volume effect can be expressed as a proper-time integral
\begin{eqnarray}
R(x_\perp)
&=&
{-}
\int_0^\infty \frac{ds}{(4 \p s F_\p)^2} \, e^{- s m_\p^2}
\Bigg[
\frac{1}{2} 
\Big(
\Theta_0 (s)
{-}\,1
\Big)
\notag \\
&& \phantom{space}
+
\frac{QB s}{\sinh QB s}
\Big( 
\Theta(x_\perp|s)
{-}\,1
\Big)
\Bigg]
.\label{eq:Rfin}
\end{eqnarray}

\begin{figure}
\begin{center}
\includegraphics[width=0.37\textwidth]{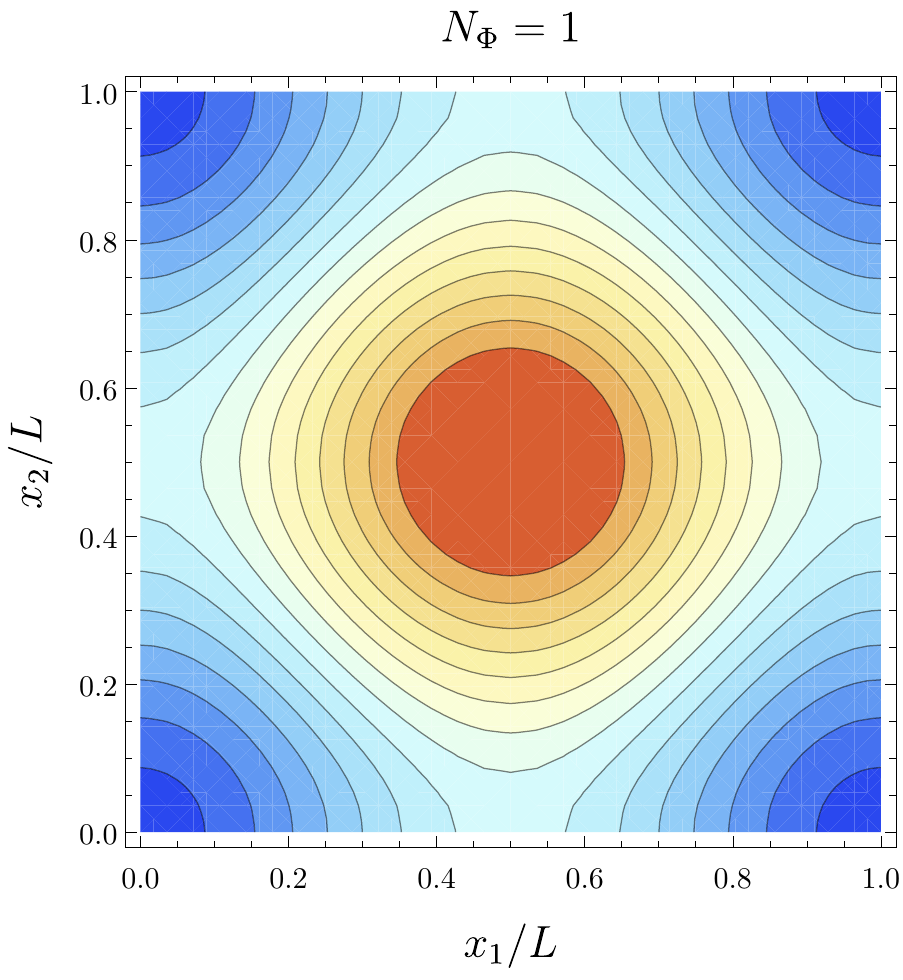}
\includegraphics[width=0.37\textwidth]{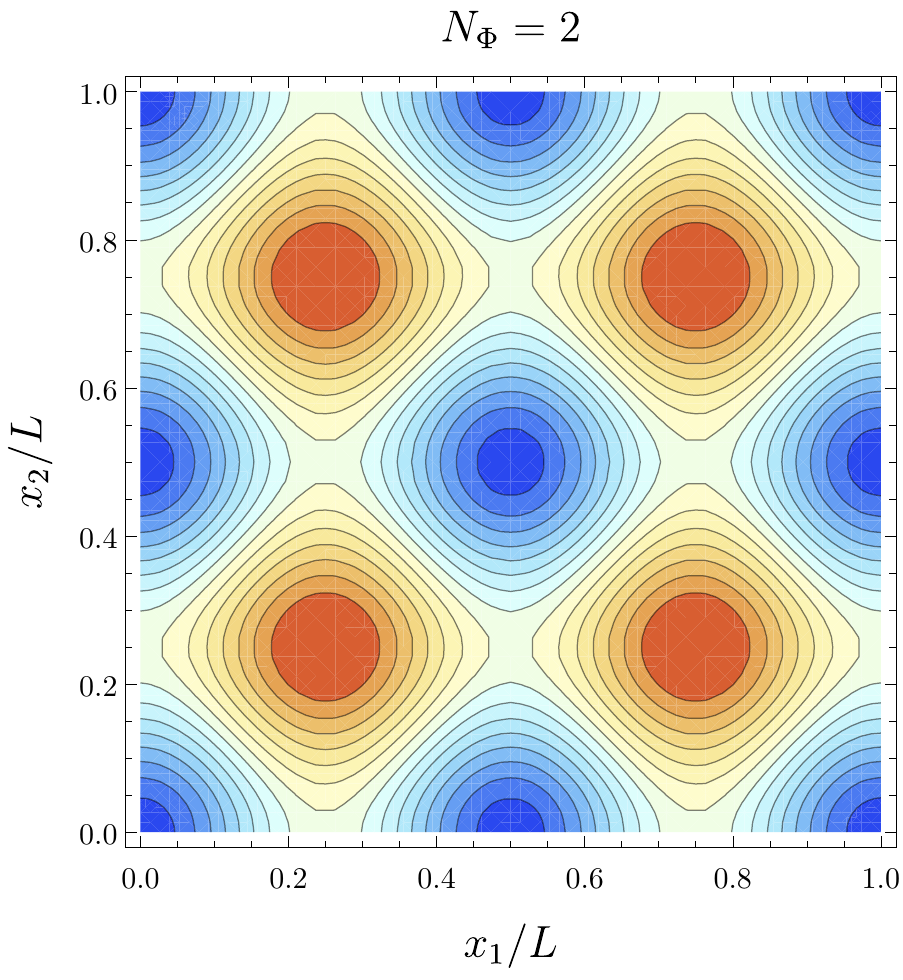}
\includegraphics[width=0.37\textwidth]{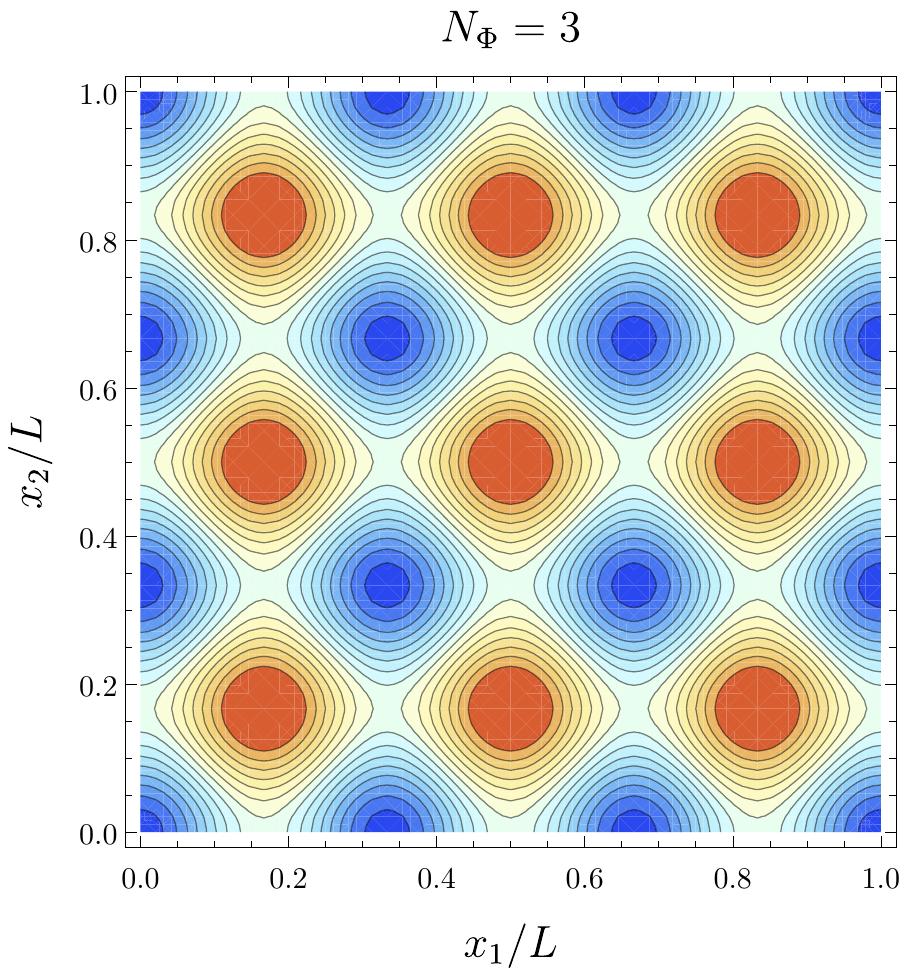}
\end{center}
\vskip-2.25em
\caption{
Contour plots of the finite volume correction to the chiral condensate. 
Heat maps of 
$R(x_\perp)$
in Eq.~\eqref{eq:qbarqLinf}
are plotted as a function of 
$x_\perp$,
with 
$m_\p L=3$
and vanishing twist angles
$\theta_\perp = 0$. 
Flux quanta 
$N_\Phi = 1$, $2$, and $3$
are displayed and exhibit single, double, and triple periodicity,
respectively. 
\color{black}
} 
\label{fig:ccosc}
\end{figure}

For nonzero values of the magnetic field, 
the finite volume effect 
$R(x_\perp)$
is coordinate dependent, 
which is depicted in Fig.~\ref{fig:ccosc}
for a cubic volume 
$V = L^3$. 
Generally the local chiral condensate has oscillatory behavior. 
In asymptotically large volumes,%
\footnote{ 
Technically only asymptotically large areas 
$A_\perp = L_1 L_2$ 
transverse to the magnetic field
are required.} 
these oscillations are exponentially suppressed. 
This suppression can be exhibited analytically by retaining the first image corrections, 
namely those that satisfy
$| \vec{\n} \,| = 1$. 
In a cubic volume, 
the leading asymptotic behavior of 
Eq.~\eqref{eq:Rfin}
takes the form
\begin{eqnarray}
R(x_\perp) 
&=&
- \left[ \frac{5}{2} + \cos \left(2 \p N_\Phi \frac{x_1}{L} \right) +  \cos \left(2 \p N_\Phi \frac{x_2}{ L} \right)  \right]
\notag \\
&& \phantom{space}
\times
\frac{m_\p^2}{F_\p^2} \,
\frac{e^{- m_\p L}}{(2 \p \, m_\p L)^{3/2}}
+ \cdots
\label{eq:qbarqLinf}
,\end{eqnarray}
where, 
for simplicity, 
we have taken vanishing twist angles. 
The large-volume limit, 
furthermore, 
is taken above as 
$m_\pi L \gg 1$
but with 
$N_\Phi$
held fixed, 
for which
$Q B / m_\p^2 \ll 1$.

Oscillations of the local condensate are completely an artifact of the finite volume. 
In lattice QCD calculations, 
moreover,
the local chiral condensate is rarely obtained, because it is statistically noisy. 
Instead, 
a volume average greatly improves the signal. 
Volume averaging the result from chiral perturbation theory has the effect of removing the oscillatory terms. 
This can already be anticipated from the simple oscillations exhibited in the asymptotic formula.
When one averages 
 Eq.~\eqref{eq:qbarqLinf}
over the transverse plane, 
the volume effect is reduced by a factor of 
$5/9$
compared to the case of vanishing magnetic field.

In the general case of 
Eq.~\eqref{eq:Rperp1loop}, 
averaging over the transverse plane projects onto the sector of zero winding numbers  
\begin{eqnarray}
\frac{1}{L_1}
\int_{0}^{L_{1}}dx_{1}\, \left[ W_{2}^{*}(x_{1}) \right]^{\nu_{1}} &=& \delta_{\nu_{1},0} 
,\notag
\\
\label{eq:winding2}
\frac{1}{L_2}
\int_{0}^{L_{2}}dx_{2}\, \left[ W_{1}^{*}(x_{2}) \right]^{\nu_{2}} &=& \delta_{\nu_{2},0} 
.\end{eqnarray}
With 
$\n_\perp = (0,0)$, 
the effect of the finite transverse area 
$A_\perp = L_1 L_2$
enters only through the magnetic field quantization condition, 
and not from transverse images. 
Due to the remnant
$\mathbb{Z}_{N_\Phi}$
translational invariance, 
volume averaging the charged pion ring diagram thus has the effect of sending
$A_\perp \to \infty$
with 
$Q B$
held fixed.

The finite volume effect on the 
volume averaged condensate
$\langle \, \ol \psi \psi \, \rangle$
can be obtained simply by performing the spacetime average. 
Using the ratio
$R(x_\perp)$
in 
Eq.~\eqref{eq:Ratio}, 
we denote the result of volume averaging by
\begin{equation}
\langle R \, \rangle
= 
\frac{1}{\be V}
\int d^4 x
\,\, R(x_\perp)
.\end{equation}
Carrying out the spacetime integral of the one-loop result
Eq.~\eqref{eq:Rfin} 
using 
Eq.~\eqref{eq:winding2}, 
we obtain
\begin{eqnarray}
\langle R\, \rangle
&=&
{-} 
\int_0^\infty \frac{ds}{(4\p s F_\p)^2} e^{- s m_\p^2} 
\Bigg[ \frac{1}{2} \Big( \Theta_0(s) {-}\,1 \Big)
\notag \\
&& \phantom{space}
+ \frac{Q B s}{\sinh Q B s} \Big( \vartheta_3 \big(0,e^{-\frac{L_3^2}{4s}} \big) {-}\, 1 \Big) \Bigg]
.\label{eq:DRbar}
\end{eqnarray}
Notice that compared to 
Eq.~\eqref{eq:Rfin}, 
there are now only magnetic field dependent images from the 
$\hat{x}_3$-direction;
this is the 
$A_\perp \to \infty$ 
limit of the charged pion contribution with 
$QB$ 
fixed.%
\footnote{
As our wording indicates throughout, 
the nature of these limits in finite volume can be delicate. 
To isolate the magnetic field dependence of the finite volume effect, 
for example,
one might subtract the zero-field limit. 
The na\"ive limit, 
however, 
does not commute with spacetime averaging
\begin{equation}
\langle R \, \rangle\big|_{B=0}
\neq 
\tfrac{1}{\be V} \int d^4 x \,
R(x_\perp) \big|_{B=0} 
\notag
.\end{equation}
A continuous zero-field limit is only obtained with a concomitant infinite transverse-area limit due to magnetic flux quantization. 
}

\begin{figure}
\begin{flushleft}
\includegraphics[width=0.475\textwidth]{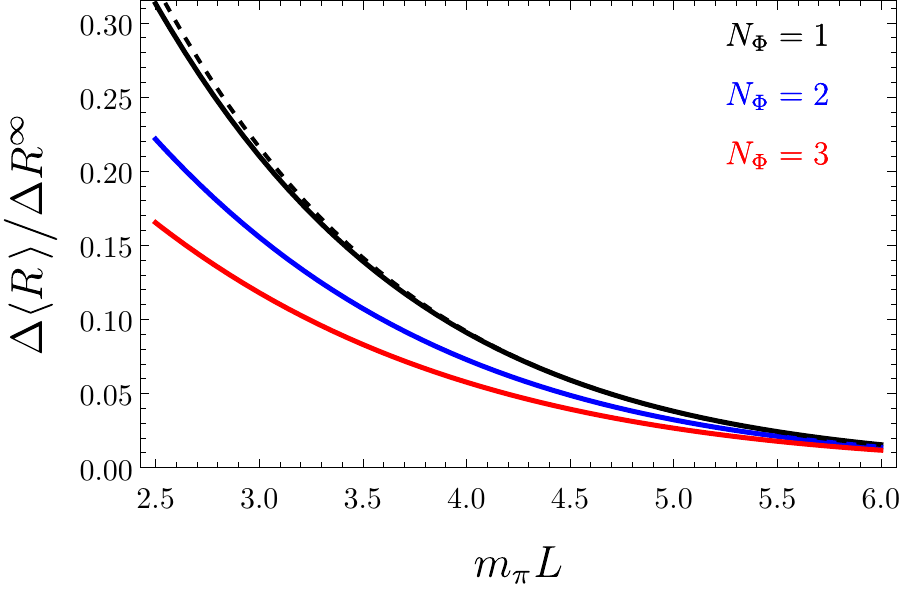}
\end{flushleft}
\vskip-1.5em
\caption{
Magnetic field dependence of the finite volume effect on the chiral condensate. 
The ratio of condensate differences in 
Eq.~\eqref{eq:D/D}
is plotted as a function of 
$m_\p L$
for different flux quanta 
$N_\Phi$. 
The leading behavior in asymptotically large volumes 
Eq.~\eqref{eq:Rasmp}
is shown as a dashed curve, 
and is independent of 
$N_\Phi$. 
The effect of finite volume is to further catalyze chiral symmetry breaking, 
but decreases with increasing flux quantum. 
} 
\color{black}
\label{fig:ccavg}
\end{figure}

The size of 
$\langle R \, \rangle$
is generally quite small, 
which means that the finite volume corrections in a magnetic field are quite small compared to 
$\langle \, \ol \psi \psi \, \rangle_0$. 
A more apt comparison, 
however,
is made by forming the ratio of differences
\begin{equation}
\frac{\D \langle R \, \rangle}{\D R^\infty} 
= 
\frac{\langle R \, \rangle - \langle R \, \rangle \big|_{B = 0}}
{\,\, R^\infty - R^\infty \big|_{B=0}} 
\label{eq:D/D}
\,\, ,\end{equation}
where
$R^\infty \equiv \langle \, \ol \psi \psi \, \rangle^\infty / \langle \, \ol \psi \psi \, \rangle_0$.
The ratio in 
Eq.~\eqref{eq:D/D}
compares the magnetic field dependence of the finite volume effect with the corresponding infinite-volume field dependence. 
The latter is magnetic catalysis of chiral symmetry breaking from chiral perturbation theory%
~\cite{Cohen:2007bt}.
In Fig.~\ref{fig:ccavg}, 
we plot the ratio of condensate differences for the case of a cubic volume
$V = L^3$.
The plot shows smaller finite-size effects for larger flux quanta. 
Increasing the flux quantum produces more compact Landau levels, 
which are naturally less sensitive to volume effects. 
Additionally, 
the finite volume effect is shown to further catalyze chiral symmetry breaking.
This is unlike finite-volume melting of the chiral condensate in a vanishing magnetic field%
~\cite{Gasser:1987ah}. 
Decreasing 
$m_\p L$
at fixed flux quantum, 
however, 
corresponds to increasing the magnetic field. 
At sufficiently small 
$m_\p L$, 
the magnetic fields become non-perturbative
$Q B / (4 \p F_\p)^2 > 1$.
Small-volume results in zero and nonzero magnetic flux are not continuously connected.

The asymptotic-volume limit of the ratio of condensate differences in 
Eq.~\eqref{eq:D/D} 
can be taken. 
Carrying out the 
$m_\p L \gg 1$ 
limit in a cubic volume at fixed 
$N_\Phi$, 
we find the flux-independent behavior
\begin{eqnarray}
\frac{\D \langle R \, \rangle}{\D R^\infty} 
&=&
\sqrt{2 \p \, m_\p L \ }
\ e^{- m_\p L}
\notag \\
&& \times
\left[ 
1 + \c O \big( \tfrac{1}{m_\p L} \big) + \c O  \big( \tfrac{N_\Phi^2}{(m_\pi L)^2}  \big) + \c O (e^{- m_\p L}) 
\right]
,\notag \\
\label{eq:Rasmp}
\end{eqnarray}
where, 
in addition to the leading power-law and exponential corrections,  
the size of the leading flux-dependent correction has been indicated. 
Fig.~\ref{fig:ccavg} 
confirms the 
$N_\Phi$-independent asymptotic behavior of the finite volume effect. 
Agreement is best for the smallest flux quantum, 
which is consistent with the scaling of flux-dependent corrections. 
For 
$N_\Phi = 1$, 
the asymptotic formula is seen to work remarkably well already at moderately large volumes
$m_\p L = 2.5$.
Sub-leading power-law corrections to the leading exponential behavior, 
however, 
are proportional to 
$(m_\p L)^{-1}$
and spoil the otherwise fortuitous agreement in moderate volumes.

\subsection{Free Energy}

The free energy density 
$\c F$
is related to the thermodynamic partition function through the relation
\begin{equation}
\c F = - \frac{1}{\be V} \log Z
.\end{equation} 
In chiral perturbation theory, 
the free energy density of pions can be obtained from 
Eq.~\eqref{eq:psibarpsi}
through the relation
\begin{equation}
\frac{\partial \c F}{\partial m^2} = F^2 \frac{\langle \, \ol \psi \psi \, \rangle}{\,\,\langle \, \ol \psi \psi \, \rangle_0}
,\label{eq:FGMOR}
\end{equation}
where
$m^2$
is the tree-level pion mass,
and the Gell-Mann--Oakes--Renner relation has been applied.  
In what follows, 
we focus on the magnetic free energy density, 
which we define as 
\begin{equation}
\c F_B = \c F - \c F\, \big|_{B=0} - \tfrac{1}{2} B^2
\label{eq:Fsubtract}
.\end{equation}
The subtraction of 
$\frac{1}{2} B^2$
removes the pure gauge contribution from the free energy density, 
so that 
$\c F_B$
is exclusively the matter contribution.

Integrating the regularized expression for the charged pion contribution to the condensate 
$\langle \, \ol \psi \psi \, \rangle$
leads to the free energy density
$\c F_B$
up to an integration constant. 
The result, 
however, 
requires renormalization. 
As in Sec.~\ref{s:IIC}, 
only the 
$\n_\m = 0_\m$
term is divergent,
and this term is the infinite-volume limit.
Hence, 
we make the further separation
\begin{equation}
\c F_B
 = 
\c F_B^\infty + \c F^\text{FV}_B
\label{eq:FB}
,\end{equation}
in order to isolate the divergent term
$\c F_B^\infty$
in the sum over images. 
From 
Eq.~\eqref{eq:FGMOR},
it has the form
\begin{equation}
\c F_B^\infty 
= 
- \int dm^2 \,
\left( 
g_+(0_\m) 
- 
g_+(0_\m)\Big|_{B=0}
\right) 
+
c_B(s_0)
,\end{equation}
where
$g_+(0_\m)$
appears in 
Eq.~\eqref{eq:offending},
and
$c_B(s_0)$
is independent of
$m^2$. 
Even after subtraction of the 
$B = 0$
limit, 
the
$\c O(B^2)$ 
term in 
$\c F_B^\infty$
logarithmically diverges
as 
$s_0 \to 0$. 
In this case, 
one requires a magnetic field-dependent counterterm from the chiral Lagrangian
$c_B \propto B^2$, 
which, 
due to its independence from pion fields, is called a high-energy constant in the language of%
~\cite{Gasser:1983yg}.
The renormalization condition on the 
$\c O(B^2)$
term of the free energy density
$\c F$
is chosen to preserve the classical value
$\frac{1}{2} B^2$, 
which ensures that the magnetic field strength is not renormalized from matter fields%
~\cite{Schwinger:1951nm}. 
After renormalization, 
the infinite volume contribution to the magnetic free energy density is thus
\begin{eqnarray}
\c F_B^\infty
&=&
\int_0^\infty
\frac{ds}{(4\pi)^{2} s^{3}} \, e^{-s m_\p^2}
\left[\frac{QBs}{\sinh QBs}-1+\frac{(QBs)^2}{6}\right]
\label{eq:Finf}
.\notag \\
\end{eqnarray}
The finite volume contribution, 
by contrast, 
requires no renormalization, 
and is given by
\begin{eqnarray}
\noindent
\c F^\text{FV}_B
&=&
\int_0^\infty \frac{ds}{(4\pi)^2 s^3} \, e^{-s m_\p^2}
\left[\frac{QBs}{\sinh QBs}-1\right]
\notag \\
&& \phantom{spaces}
\times
\left[\vartheta_3\big(0,e^{-\frac{L_3^2}{4s}}\big)-1\right]
\label{eq:Ffin}
.\end{eqnarray}
In Eqs.~\eqref{eq:Finf} and \eqref{eq:Ffin}, 
we have replaced the tree-level pion mass 
$m$ 
with the physical pion mass 
$m_\p$, 
because the difference is higher order.
Notice that the finite volume effect involves only images in the longitudinal direction. 
This is the 
$A_\perp \to \infty$
limit with 
$QB$
fixed that we encounter above for the chiral condensate.
The free energy is a spacetime average, 
rather than a local distribution.

\subsection{Magnetic Pressure Anisotropy and Magnetization}
\label{s:aniso}

The addition of a magnetic field breaks the isotropy of space, 
which can manifest itself in thermodynamic quantities. 
Allowing for such anisotropy, 
the matter contribution to the magnetic pressure in the 
$i^{\rm th}$ 
direction can be defined as%
~\cite{Ferrer:2010wz}
\begin{eqnarray}
p_i
&=&
- \frac{L_{i}}{V} \left(\frac{\partial F_B}{\partial L_{i}}\right)_{L_{j},B}
\label{eq:pfixB}
,\end{eqnarray}
where 
$j {\neq} i$
and 
$F_B$
is the free energy, 
namely
$F_B = V \c F_B$. 
In this definition, 
the magnetic field is held fixed. 
Due to the field quantization condition on a torus, 
however, 
lattice practitioners cannot generally vary the size of a transverse direction while keeping the magnetic field fixed.
In lattice QCD calculations, 
a more accessible definition of the magnetic pressure is%
~\cite{Bali:2013esa}
\begin{eqnarray}
\widetilde{p}_i
&=&
- \frac{L_{i}}{V} \left(\frac{\partial F_B}{\partial L_{i}}\right)_{L_{j},N_{\Phi}}
\label{eq:pfixN},
\end{eqnarray}
which is defined at fixed magnetic flux. 
In the longitudinal direction, 
these two definitions of pressure are identical
\begin{equation}
\widetilde{p}_{3}
=
p_{3}
=
-\c F_B 
- L_3 \frac{\partial \c F_B}{\partial L_3}
.\end{equation}
Due to magnetic flux quantization, 
however,
the magnetic pressure in the transverse directions differs between the two definitions
\begin{equation}
\widetilde{p}_{\perp}
-
p_{\perp}
=
B
\left(\frac{\partial\c F_B}{\partial B} \right)_V
\equiv
-
B
\mathcal{M}_{r}
\label{eq:Dpperps}
,\end{equation}
where we have identified the renormalized magnetization 
$\c M_r$, 
given by 
\begin{equation}
\c M_r 
= 
- \left( \frac{\partial \c F_{B}}{\partial B} \right)_V
\label{eq:Mr}
.\end{equation} 
The renormalized magnetization is a measure of the magnetic response of the 
QCD vacuum%
~\cite{Kabat:2002er,Cohen:2008bk}. 
Note that by subtracting the gauge contribution to the free energy density 
$\c F_B$
in Eq.~\eqref{eq:Fsubtract}, 
$\c M_r$
depends only on the matter contribution to the free energy density.

To access the renormalized magnetization in lattice QCD calculations, 
one can measure the magnetic pressure anisotropy%
~\cite{Bali:2013esa}
\begin{equation}
\D \widetilde{p}
\equiv
\widetilde{p}_{\perp}-\widetilde{p}_{3}
.\end{equation}
Using 
Eqs.~\eqref{eq:pfixN}
and 
\eqref{eq:Mr}, 
we arrive at the expression for the matter contribution to the pressure anisotropy
\begin{equation}
\D \widetilde{p}
=
-B \c M_r 
+ 
L_3 \frac{\partial \c F_B}{\partial L_3}
\label{eq:dp}
.\end{equation}
The second term in the pressure anisotropy is solely an artifact of the finite size of the longitudinal direction.
Taking 
$L_3 \to \infty$
with 
$QB$
held fixed, 
we have
\begin{equation}
\D \widetilde{p} \, {}^\infty
=
-B \c M_r^\infty 
\label{eq:dpinf}
,\end{equation}
where the infinite-volume limit of the magnetization 
$\c M_r^\infty$
arises from the infinite-volume limit of the free energy density
$\c M_r^\infty = - \left( \partial \c F_B^\infty / \partial B \right)_V$.

\begin{figure}
\begin{flushleft}
\includegraphics[width=0.47\textwidth]{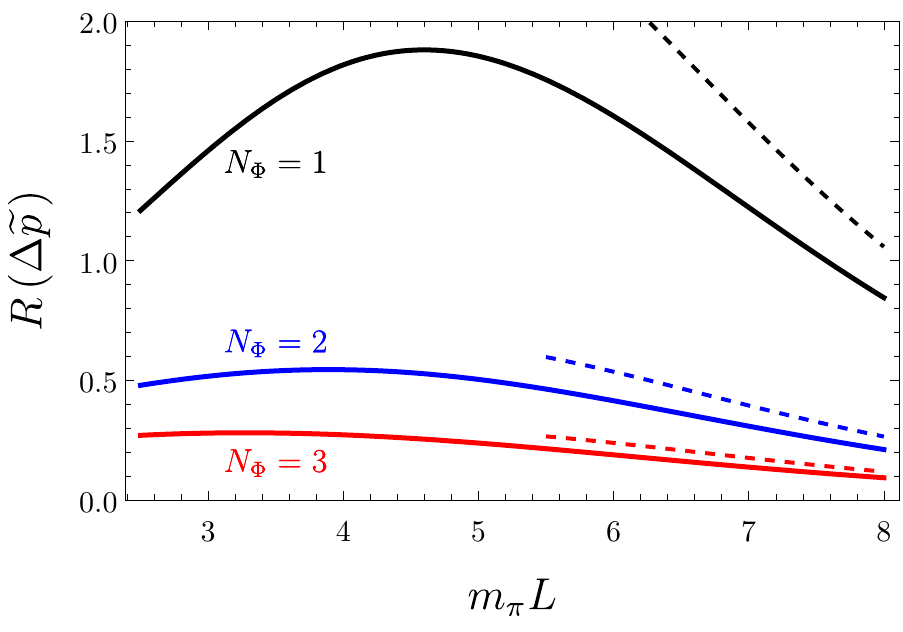}
\end{flushleft}
\vskip-2em
\caption{
Finite volume effect on the magnetic pressure anisotropy. 
Plotted as a function of 
$m_\p L$
is the ratio
$R\left(\D \widetilde{p} \,\right)$
in 
Eq.~\eqref{eq:dpRatio}
of the finite volume effect on the magnetic pressure anisotropy 
$\D \widetilde{p}$
compared to its infinite volume value. 
Results are plotted for the lowest three flux quanta, 
with the corresponding large-volume behavior 
Eq.~\eqref{eq:Rp_asymp}
shown as dashed curves.  
Finite volume effects persist to large values of 
$m_\p L$, 
but these correspond to small values of the magnetic field, 
where the renormalized magnetization 
$\c M_r^\infty$
is itself small.
}
\label{fig:anisotropy}
\end{figure}

In Fig.~\ref{fig:anisotropy}, 
we investigate the finite volume correction to the magnetic pressure anisotropy 
for a cubic volume 
$V = L^3$. 
The ratio 
$R\left(\D \widetilde p\,\right)$
of the finite volume effect compared to the infinite volume anisotropy
\begin{equation}
R\left(\D \widetilde p \, \right)
=
\frac{\D \widetilde{p} - \D \widetilde{p} \, {}^\infty}{\D \widetilde{p} \, {}^\infty}
\label{eq:dpRatio}
,\end{equation} is plotted 
as a function of 
$m_\p L$
for the lowest values of the flux quantum
$N_\Phi$. 
The finite volume effect generally decreases with increasing flux quantum. 
This is sensible as larger magnetic fields on a fixed-sized lattice lead to more compact Landau levels, 
which are naturally less sensitive to the finite volume. 
The effect of the finite volume, 
however, 
is shown to be substantial in the figure. 
This we anticipate because both the infinite-volume and finite-volume results are loop effects of the same order. 
Additionally, 
a vanishing finite volume effect in the  
$m_\p L \to \infty$
is very slowly obtained. 
This limit does not correspond to a fixed magnetic field. 
Instead, 
the magnetic field strength decreases with increasing 
$m_\p L$. 
The renormalized magnetization
$\c M_r^\infty$
is itself small when the field is small, 
namely 
$\c M_r^\infty \propto B^3$. 
Combined with the asymptotic expansion of the integrand of Eq.~\eqref{eq:dpRatio}, 
we obtain the large-volume behavior 
\begin{eqnarray}
R\left(\D \widetilde p \, \right)
&=& 
\frac{30}{7(2 \p)^{3/2} N_\Phi^2}
\, (m_\p L)^{9/2} \, e^{ - m_\p L}
\notag \\
&& \phantom{space}
\times
\left[ 1 + \c O \left(\tfrac{1}{m_\p L} \right)
+ 
\c O \left(e^{ - m_\p L} \right)
\right]
,\quad\,\,
\label{eq:Rp_asymp}
\end{eqnarray}
taken at fixed 
$N_\Phi$.
This leading exponential behavior arises exclusively from 
$L_3 \big( \partial \c F_B^\text{FV} / \partial L_3 \big)_B$, 
with contributions from 
$B \big( \partial \c F_B^\text{FV} / \partial B  \big)_V$
suppressed by a relative factor of
$(m_\p L)^{-1}$. 
Fig.~\ref{fig:anisotropy}
confirms that this asymptotic behavior does not set in until rather large values of 
$m_\p L$.

\section{Neutral Pion Effective Action}
\label{sec:action}

Beyond thermodynamic quantities, 
hadron energies and interactions are modified in background fields. 
The simplest hadron to consider is the neutral pion, 
which feels the effect of the magnetic field through virtual charged pion fluctuations. 
The effective action for the neutral pion in a magnetic field is determined
at one-loop order in the chiral expansion in Sec.~\ref{s:Comp}. 
In finite volume, 
the effective action for a neutral particle has explicit coordinate dependence due to the remnant 
$\mathbb{Z}_{N_\Phi}$
translational invariance.%
\footnote{ 
An analogous result was obtained earlier for the case of the neutron effective action 
calculated using heavy baryon chiral perturbation theory%
~\cite{Tiburzi:2014zva}.
} 
The effective action obtained is then utilized in 
Sec.~\ref{s:Two} 
to compute the two-point correlation function of the 
neutral pion. 
While the general result is quite complicated, 
averaging over the source location leads to dramatic simplifications due to charge neutrality.

\subsection{Computation of the Effective Action}
\label{s:Comp}

\subsubsection{One-Loop Computation}

At next-to-leading order, 
the four pion terms of the Euclidean action density in two-flavor chiral perturbation theory are%
~\cite{Gasser:1983yg}
\begin{eqnarray}
\c L
&=&
{-}
\tfrac{m^2}{24 F^2} (\pi^0)^4 
{-}
\tfrac{m^2}{6 F^2} \pi^+ \pi^- (\pi^0)^2
{-}
\tfrac{1}{3F^2} \pi^+ \pi^- (\partial_\mu \pi^0)^2
\notag \\
&& 
{-}
\tfrac{1}{3F^2} D_\mu \pi^+ D_\mu \pi^- (\pi^0)^2
{+} 
\tfrac{1}{3F^2} \partial_\mu (\pi^+ \pi^-) \pi^0 \partial_\mu \pi^0
.\notag\\
\label{eq:NLO}
\end{eqnarray}
These terms produce a perturbative correction to the neutral pion propagator 
$G_0(x',x)$
of the form
\begin{eqnarray}
\d G_0(x',x)
=
\int_y G_0(x',y) \, \c O(y) \, G_0(y,x)
\label{eq:dG0}
,\end{eqnarray}
where we use 
$\int_y$
as an abbreviation for the integral 
$\int d^4 y$
over finite spacetime.
From Eq.~\eqref{eq:NLO}, 
the coordinate-space operator
$\c O(y)$
arising from the one-loop tadpole diagrams has the form
\begin{eqnarray}
\c O(y)
&=&
\overleftarrow{\partial_\mu}  V_1(y)  \overrightarrow{\partial_\mu}
-
V_2(y)
\notag \\
&&
-
\tfrac{1}{2}
\overleftarrow{\partial_\mu}
[\partial_\m  V_1(y)]
-
\tfrac{1}{2}
[\partial_\m  V_1(y)] 
\overrightarrow{\partial_\mu}
,\label{eq:O(y)}
\end{eqnarray}
where the partial derivatives are all taken with respect to 
$y$. 
In writing this perturbative correction, 
we have defined 
\begin{eqnarray}
V_1(y) 
&=&
\tfrac{2}{3F^2} G_+(y,y),
\notag \\
V_2(y)
&=&
- \tfrac{m^2}{2F^2} G_0(0,0) + \tfrac{m^2}{2} V_1(y) - \tfrac{1}{2} [\partial^2_\m V_1(y)]
.\end{eqnarray}
The neutral pion propagator is translationally invariant;
and, 
we have accordingly written 
$G_0(y,y) = G_0(0,0)$, 
as in 
Sec.~\ref{s:IIC}.
There is a term contributing to 
$V_2(y)$ 
that contains two covariant derivatives of the charged pion propagator, 
and this has been simplified using the relation in
Eq.~(\ref{eq:DDG}).

To further simplify 
$\c O(y)$, 
it is efficacious to write the first term in 
Eq.~\eqref{eq:O(y)}
as half the sum of two terms, 
where each term is the result of one of the two possible integrations by parts.
The boundary terms produced vanish due to periodicity.%
\footnote{
After integration by parts on the variable
$y_\mu$, 
the two boundary terms produced are 
$G_{0}(x',y)G_{+}(y,y)\partial_{\mu}G_{0}(y,x) \big|_{y_\mu = 0}^{y_\mu = L_\mu}$
and
$\partial_{\mu}G_{0}(x',y)G_{+}(y,y)G_{0}(y,x) \big|_{y_\mu = 0}^{y_\mu = L_\mu}$, 
up to constants of proportionality.  
Both of these terms vanish, which is due to periodicity of the neutral pion propagator, 
and periodicity of $G_+(y,y)$. 
The latter owes to Eq.~\eqref{eq:fperp} and the periodicity of the Wilson lines in Eq.~\eqref{eq:DTI}, 
\emph{i.e.}, 
they are also invariant for 
$n = N_\Phi \notin \mathbb{Z}_{N_\Phi}$. 
}
Carrying out integration by parts symmetrically thus enables the replacement
\begin{eqnarray}
\overleftarrow{\partial_{\mu}}V_{1}(y)\overrightarrow{\partial_{\mu}}
&=&
-
\tfrac{1}{2}
\left(
\overleftarrow{\partial_{\mu}}
V_{1}(y)
\overleftarrow{\partial_{\mu}}
+
\overrightarrow{\partial_{\mu}}
V_{1}(y)
\overrightarrow{\partial_{\mu}}
\right)
.\end{eqnarray}
With this replacement made, 
the coordinate-space operator 
$\c O(y)$ in 
Eq.~\eqref{eq:O(y)}
subsequently becomes
\begin{eqnarray}
\c O(y)
&=&
-
\tfrac{1}{2}
\left(
\overleftarrow{\partial_{\mu}}
\overleftarrow{\partial_{\mu}}
V_{1}(y)
+
V_{1}(y)
\overrightarrow{\partial_{\mu}}
\overrightarrow{\partial_{\mu}}
\right)
- 
V_2(y)
\notag \\
&&-
\overleftarrow{\partial_{\mu}}[\partial_{\mu}V_{1}(y)]
-
[\partial_{\mu}V_{1}(y)]\overrightarrow{\partial_{\mu}}
.\end{eqnarray}
The first two terms above will permit simplification in 
Eq.~\eqref{eq:dG0} 
using the Green's function relation for the neutral pion.
The last two terms above simplify when one of them is integrated by parts, 
with the boundary contribution again vanishing  due to periodicity.  
The net result of these manipulations is the simplified form
\begin{eqnarray}
\mathcal{O}(y)
&=&
\left(
- 
\overleftarrow{\partial_{\mu}}
\overleftarrow{\partial_{\mu}}
+
m^2
\right)
\tfrac{1}{2}
V_{1}(y)
\notag \\
&&
+
\tfrac{1}{2}
V_{1}(y)
\left(
- 
\overrightarrow{\partial_{\mu}}
\overrightarrow{\partial_{\mu}}
+
m^2
\right)
- 
V(y)
,\label{eq:Ofinal}
\end{eqnarray}
where
\begin{eqnarray}
V(y)
&=&
m^2
V_{1}(y)
+
V_{2}(y)
-
\partial^2_\m V_{1}(y)
.\label{eq:V}
\end{eqnarray}

The coordinate-space operator 
$\c O(y)$
can be viewed as the action of an abstract operator  
$\c O$
in coordinate space, 
namely
$| y \rangle \c O(y) \langle y | =  \c O | y \rangle\langle y |$. 
Written in this way, 
the Green's function itself is an operator. 
We denote the free Green's function operator as 
$G_0$,
and the full Green's function operator as 
$\c G_0$. 
Including the perturbative correction in Eq.~\eqref{eq:dG0}
with 
$\c O$
from Eq.~\eqref{eq:Ofinal}, 
the Green's function operator
$\c G_0$
is thus of the form 
\begin{eqnarray}
\c G_0
&=&
G_0
+
\tfrac{1}{2} \left( V_1 G_0 + G_0 V_1 \right) 
- 
G_0  V  G_0
.\end{eqnarray}
The effective action is the inverse of the operator
$\c G_0$, 
which at next-to-leading order accuracy is given by
\begin{eqnarray}
\c G_0^{-1}
&=&
(1 - \tfrac{1}{2} V_1) G_0^{-1} ( 1 - \tfrac{1}{2} V_1)
+
V
+ 
\cdots
.\end{eqnarray}   
Wavefunction renormalization can be accomplished by 
employing the coordinate-dependent field redefinition 
\begin{equation}
\pi^0 
\longrightarrow 
\widetilde{\pi}^0
= 
\left( 1 + \tfrac{1}{2} V_1 \right) 
\pi^0
.\end{equation} 
After this field redefinition, 
we arrive at a neutral pion effective action with the canonical normalization
\begin{eqnarray}
\widetilde{\c G}_0^{-1}
&=&
G_0^{-1} 
+
V
+ 
\cdots
,\end{eqnarray}   
where 
$V$
in 
Eq.~\eqref{eq:V}
is now identified as the effective potential. 
It is given in coordinate space in terms of coincident propagators as
\begin{equation}
V(y)
=
\tfrac{m^2}{F^2}
\left[
G_+(y,y)
{-} 
\tfrac{1}{2} G_0(0,0)
\right]
{-}
\tfrac{1}{F^2} \partial^2_\m G_+(y,y)
\label{eq:V(y)}
.\end{equation}
For ease of notation, 
the tilde will subsequently be dropped from the redefined neutral pion field.

\subsubsection{Renormalization}
\label{s:renorm}

The expressions written above have an ultraviolet divergence in 
$V$
that is regulated by a proper-time cutoff 
$s_0 \ll 1$.
As in Sec.~\ref{s:IIC},  
the divergence is independent of both the magnetic field and the volume, 
and is canceled by inclusion of the appropriate counterterms from the chiral Lagrangian. 
A further effect of such counterterms is a shift of the mass-squared appearing in 
$G_0^{-1}$. 
After the divergence is canceled, 
one can renormalize the tree-level mass 
$m$ 
to the physical pion mass 
$m_\pi$, 
\emph{i.e.}~the mass in the zero-field and infinite-volume limit. 
There is no remaining 
$s_0$
dependence in this renormalization scheme.

To focus specifically on finite volume effects, 
we arrange the terms of the effective action by adding and subtracting the infinite-volume limit in nonzero magnetic fields.
To this end, 
the renormalized neutral pion effective action is written as
\begin{equation}
\c L_\text{eff}
=
\tfrac{1}{2} 
\left(\partial_\mu \pi^0\right)^2
+
\tfrac{1}{2} 
\left( E^2 + \c V \, \right)
\left(\pi^0 \right)^2
\label{eq:EffA2}
.\end{equation}
The neutral pion energy 
$E$
is defined from the relation 
\begin{eqnarray}
E^2 
&=&
m_\pi^2 
\left[ 
1
+
\frac{Q B}{(4 \p F_\p)^2} 
\,
\c I \left( \frac{m_\pi^2}{Q B} \right)
\right]
\label{eq:E}
,\end{eqnarray}
where the magnetic field dependence arises through the function
$\c I (\a)$
given in Eq.~\eqref{eq:Ialpha}. 
This infinite-volume result was obtained using chiral perturbation theory in 
Ref.~\cite{Tiburzi:2008ma}.

The remaining contribution to the effective action appearing in  
Eq.~\eqref{eq:EffA2} 
is the effective potential
$\c V = \c V(y_\perp)$.
This is defined to be solely a finite volume effect,%
\footnote{
As such, 
the volume effect includes that in a vanishing magnetic field. 
Setting
$Q = 0$, 
we obtain
$\c V \, \big|_{Q=0} = \D m_\p^2$, 
which is the $p$-regime finite volume effect on the pion mass-squared%
~\cite{Gasser:1987zq}.
Taking the magnetic field to vanish, 
we obtain 
$\c V \, \big|_{B=0} = \D m_{\p^0}^2(Q \, \vec{\theta}_\perp)$, 
where 
$\D m_{\p^0}^2(\vec{\vartheta})$
is the finite volume effect on the neutral pion mass-squared in the presence of isospin twisted boundary conditions%
~\cite{Sachrajda:2004mi}, 
where the twist angles are identified as 
$\vec{\vartheta} = Q \, \vec{\theta}_\perp$.
} 
for which we have
\begin{eqnarray}
\c V(y_\perp)
=
\frac{m_\pi^2}{F_\p^2}
\left[
\ol G_+(y,y) 
-
\frac{1}{2}
\ol G_0(0,0)
\right]
-
\frac{\partial^2_\mu G_+(y,y)}{F_\p^2} 
.\notag\\
\label{eq:Vyperp}
\end{eqnarray}
Note that the contribution to the effective potential from 
$\partial^2_\m G_+(y,y)$, 
which is given in 
Eq.~\eqref{eq:ddG+},
automatically vanishes in infinite volume, 
and additionally vanishes in zero magnetic field.

\subsection{Two-Point Function Computation}
\label{s:Two}

The coordinate dependence of the effective potential
Eq.~\eqref{eq:Vyperp} 
leads to complicated behavior for the two-point 
correlation function of the neutral pion. 
We compute this behavior generally, 
but show that source location averaging produces dramatic simplifications
due to charge neutrality. 
The finite-volume correction to the neutral pion energy is straightforwardly
identified after source location averaging.

Using the renormalized effective action in 
Eq.~\eqref{eq:EffA2}
to compute the two-point function, 
we arrive at 
\begin{eqnarray}
\c G_0 (x',x)
= 
\widetilde{G}_0(x',x)
- 
\int_y 
\widetilde{G}_0(x',y) \, \c V(y_\perp) \, \widetilde{G}_0(y,x)
,\,\,\,
\label{eq:twopt}
\end{eqnarray}
where the effective potential has been treated perturbatively. 
At zero temperature, 
the free neutral pion propagator has become
\begin{equation}
\widetilde{G}_0(x',x)
=
\int_{-\infty}^\infty \frac{dp_0}{2 \pi}
\sum_{\vec{p}} 
\frac{\, e^{ i p_\mu (x' - x)_\mu }}{p_\mu p_\mu + E^2}
,\end{equation}
where the sum runs over periodic momentum modes. 
Our convention is that 
$\sum_{\vec{p}} \equiv \frac{1}{V} \sum_{\vec{n}}$, 
where 
$p_i = \frac{2\p}{L_i} n_i$
and each mode number
$n_i \in \mathbb{Z}$. 
The tilde reflects that 
$G_0$
differs from 
$\widetilde{G}_0$
by the replacement 
$m^2 \to E^2$.

In a typical lattice QCD calculation of the spectrum, 
one performs a projection onto zero spatial momentum at the sink
$x'$.  
With infinite temporal extent, 
the zero-momentum projected two-point correlation function is defined by 
\begin{eqnarray}
\c C_{0}(T,x_\perp)
&=&
\int_{\vec{x}'}
\c G_0(x',x)
,\end{eqnarray}
where 
$T > 0$ 
denotes the Euclidean time separation, 
$T \equiv x'_0 - x_0$. 
The correlation function retains dependence on the transverse location of the source
$x_\perp$, 
as the notation indicates. 
Using Eq.~\eqref{eq:twopt} to obtain the zero momentum projected two-point function, 
it can be written in the form
\begin{eqnarray}
\c C_0 (T, x_\perp)
&=& 
\frac{e^{-(E+\D E) T}}{2(E +\D E)}
\Big[ 1+ \c R_\perp(T, x_\perp, \theta_\perp) \Big]
\label{eq:G00}
,\end{eqnarray}
where the $T$-dependence of the function 
$\c R_\perp$
modifies the simple exponential falloff.     
In addition to the transverse location of the source, 
the function
$\c R_\perp$ 
also depends on the twist angles
$\theta_\perp$.
The detailed derivation of 
$\c R_\perp(T,x_\perp,\theta_\perp)$
and the energy shift
$\D E$
is given in 
App.~\ref{s:C}.

\subsubsection{Behavior of the Correlation Function}

To investigate the modification of the two-point function
$\c C_0(T,x_\perp)$
in Eq.~\eqref{eq:G00}, 
we consider the scenario of vanishing uniform gauge potential
$\theta_\perp = 0$, 
with the source location chosen to be coincident with the gauge origin
$x_\perp = 0$. 
In this scenario, 
the modification function becomes
$\c R_\perp(T,0,0)$, 
but this is not much simpler than the general form given in 
Eq.~\eqref{eq:Rmess}. 
To illustrate the complications analytically, 
we focus specifically on contributions from images with
$|\vec{\n}_\perp| = 1$, 
which produce the dominate finite volume effect near the infinite-volume limit.%
\footnote{
Images with 
$|\vec{\n}\,| = 1$
give the dominant finite volume effects;
however, 
those with 
$| \n_3| = 1$
and 
$\n_\perp = 0$
are excluded from 
$\c R_\perp(T,x_\perp,\theta_\perp)$
by Eq.~\eqref{eq:Rdecomp}, 
and are accounted for in 
$\D E$. 
} 
To further simplify, 
we take a cubic volume
$V = L^3$.

In the function 
$\c R_\perp(T,0,0)$, 
momentum is not conserved between the source and sink. 
For all four images with 
$| \vec{\n}_\perp  | = 1$, 
the contributions are from source momenta of magnitude
$p \equiv \frac{2 \p}{L} N_\Phi$. 
From such images, 
we have the contribution to  
$\c R_\perp(T,0,0)$
of the form
\begin{eqnarray}
\c R_\perp^{(1)}(T)
\,{=}\,
4
\left(
1  {+} \tfrac{m_\pi^2}{p^2}
\right)
\left(
\tfrac{m_\p}{E_p} e^{(m_\p - E_p)T}  {-} 1
\right)
\tfrac{g_+ \big( |\n_\perp| {=} 1 \big)}{F_\p^2}
,\notag \\
\label{eq:Rmess1}
\end{eqnarray}
where factors of 
$E$
have been replaced by 
$m_\p$, 
as the difference is 
$\c O(F_\p^{-4})$. 
As a result of the non-vanishing source momentum
$p$, 
this contribution exhibits  
$T$
dependence.%
\footnote{
In the regime 
$p^2 \ll m_\p^2$, 
we are close to momentum conservation between the source and sink, 
and a shift of the neutral pion energy can be identified. 
This regime, 
however,
requires prohibitively large volumes
$m_\pi L \gg 2 \pi N_\Phi$, 
especially in light of the fact that several values of the flux quantum 
are needed to investigate magnetic field dependence on a fixed-size lattice.
}

\begin{figure}
\begin{flushleft}
\includegraphics[width=0.475\textwidth]{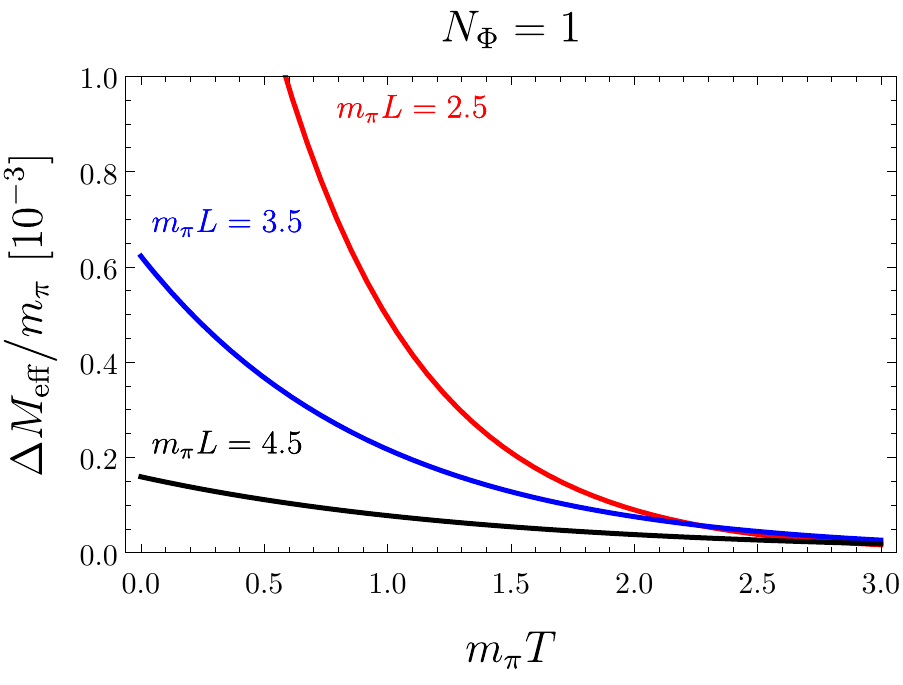}
\end{flushleft}
\vskip-2em
\caption{Finite volume effect on the effective mass function. 
Plotted versus 
$m_\p T$
is the effective mass 
$\D M_\text{eff}$ 
of the neutral pion two-point function
in units of 
$m_\p$. 
The offset in 
$\D M_\text{eff}$ 
Eq.~\eqref{eq:DT} 
ensures that the long-time plateau occurs at zero.
The effective mass of the finite-volume correlation function 
Eq.~\eqref{eq:DTR1}
only accounts for images with   
$| \n_\perp | = 1$.  
Results are plotted for 
$N_\Phi = 1$, 
for which the effect is greatest. 
} 
\label{fig:correlator}
\end{figure}

In Fig.~\ref{fig:correlator}, 
we plot the logarithmic derivative of the correlation function
\begin{equation}
\D M_\text{eff}(T) 
= 
- \frac{d}{dT} \log \, \c C_0(T,0)
- ( E + \D E)
,\label{eq:DT}
\end{equation}
using Eq.~\eqref{eq:G00}
with
$\c R_\perp(T,x_\perp,\theta_\perp)$
approximated by 
$\c R^{(1)}_\perp(T)$
given in Eq.~\eqref{eq:Rmess1}. 
In the definition of 
$\D M_\text{eff} (T)$, 
we have subtracted the long-time limit 
$E + \D E$, 
so that the effective-mass plateau will occur at zero. 
Using perturbation theory, 
we can further approximate the effective mass function as 
\begin{equation}
\D M_\text{eff} (T)
=
- \frac{d}{dT}  \c R_\perp^{(1)}(T)
\label{eq:DTR1}
,\end{equation}
up to corrections of 
$\c O(F_\p^{-4})$. 
The figure shows the finite volume effect is quite small.%
\footnote{One should keep in mind that the pion mass is not the best scale to compare the effective mass with.
To measure the magnetic polarizability of the neutral pion,
 one needs to be able to discern small shifts in the pion's energy. 
Nonetheless, 
compared to a
$\sim 10\%$ 
shift in 
$m_\pi$, 
the finite volume effect shown in 
Fig.~\ref{fig:correlator} is still small.
}
The overall size of the effect is proportional to 
$m_\p^2 / (4 \p F_\p)^2$, 
and we use the physical values 
$m_{\p} = 139.6 \ \texttt{MeV}$ and $F_{\p} = 92.1 \ \texttt{MeV}$%
~\cite{ParticleDataGroup:2022pth}.

Note that the finite-volume correction to the two-point function is not constrained by spectral positivity.
The correlation function has been projected onto zero momentum at the sink, 
and momentum is not conserved. 
The correlator cannot be written as a sum of probabilities for various contributing states; 
instead, 
there are transitions between the source and sink. 
Contributions from the transition between the lowest image momentum states and the zero momentum state, 
however,  
are positive
(as shown in Fig.~\ref{fig:correlator}), 
and appear similar to excited-state contamination;
they are exponentially suppressed for    
$(E_p - m_\p) T \gg 1$. 
Fortunately, 
we estimate this effect on the correlation function to be quite small; 
furthermore, 
it can be mitigated by source averaging.

\subsubsection{Source Location Averaging}

To cut down on statistical noise, 
the common lattice QCD procedure to compute the two-point function consists of varying the source location.%
\footnote{
Above, 
we consider averages over the source location with a fixed uniform gauge potential. 
One can also keep the source location fixed, 
and average over the twist angles; 
or even, 
vary both the source location and twist angles. 
These possibilities were discussed in the context of charged particle correlation functions in 
Ref.~\cite{Chang:2015qxa}, 
where different behaviors were contrasted. 
In the present case of the neutral pion correlation function, 
however,
all three possibilities share the salient feature that
$\c R_\perp(T,x_\perp,\theta_\perp)$
averages to zero, 
see App.~\ref{s:C}. 
}
Taking 
$N_s$
sources to be located at the transverse positions
$\{ x_{\perp 1}, x_{\perp 2}, \cdots, x_{\perp N_s} \}$, 
one can compute the source averaged correlation function
\begin{equation}
\langle \, \c C_0(T) \, \rangle_{N_s}
=
\frac{1}{N_s} \sum_{i = 1}^{N_s} \c C_0(T, x_{\perp i})
.\end{equation}
For a sufficiently large number of sources, 
one can assume that the finite source average is a reasonable approximation to the integral over all source locations. 
In this approximation, 
one has
\begin{equation}
\langle \, \c C_0(T) \, \rangle = \lim_{N_s \to \infty} \langle \, \c C_0(T) \, \rangle_{N_s}
,\end{equation}
where
\begin{equation}
\langle \, \c C_0(T) \, \rangle = \frac{1}{\be V} \int d^4 x \, \, \c C_0(T,x_\perp)
,\end{equation}
is the spacetime average of the correlation function with respect to the source location.

The spacetime average can be computed for the formula in 
Eq.~\eqref{eq:G00}. 
As detailed in App.~\ref{s:C}, 
the modification to the correlation function averages to zero
$\int d^2 x_\perp \, \c R_\perp(T,x_\perp,\theta_\perp) = 0$.  
The average over the transverse plane thus has the effect of sending
$A_\perp \to \infty$
with 
$QB$
held fixed.
Consequently, 
we recover a simple exponential falloff of the correlation function from averaging a large number of sources
\begin{eqnarray}
\langle \, \c C_0(T) \, \rangle 
= 
\frac{e^{-(E+\D E) T}}{2(E+\D E)}
\label{eq:C00}
.\end{eqnarray}
After source averaging, 
$\D E$
can be identified as the finite volume effect on the neutral pion energy. 
From App.~\ref{s:C}, 
it can be expressed by the proper-time integral
\begin{eqnarray}
\D E 
&=&
\frac{m_\p}{2}
\int_0^\infty ds \, \frac{e^{ - s m_\p^2}}{(4\p s F_\p)^2}
\Bigg[
\frac{QB s}{\sinh QB s}
\notag\\ 
&&
\times
\left( 
\vartheta_3 \big(0,e^{- \frac{L_3^2}{4s}}\big)
{-}\,1
\right)
-
\frac{1}{2} 
\Big(
\Theta_0 (s)
{-}\,1
\Big)
\Bigg]
.\label{eq:DE}
\end{eqnarray}

\begin{figure}
\begin{flushleft}
\includegraphics[width=0.475\textwidth]{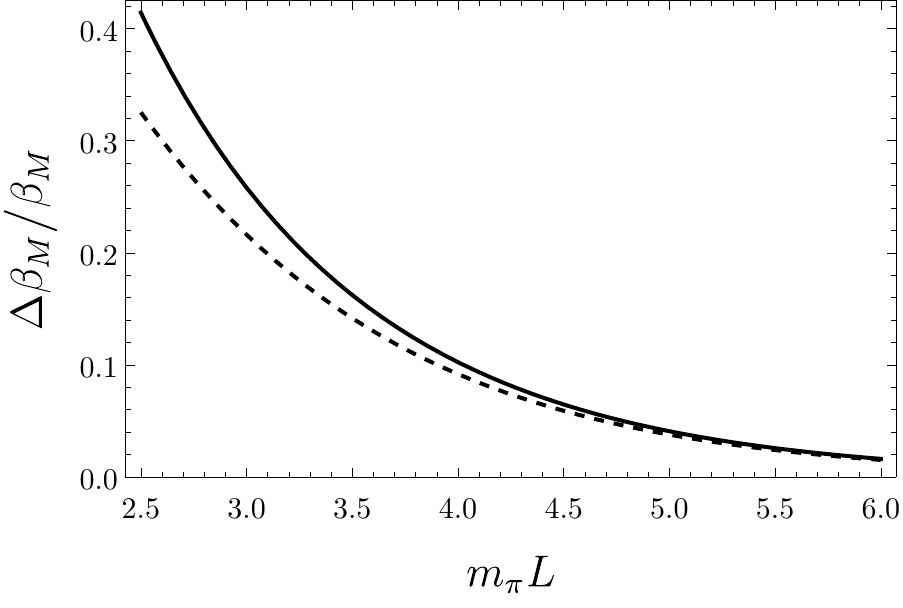}
\end{flushleft}
\vskip-2em
\caption{Finite volume effect on the magnetic polarizability of the neutral pion. 
Plotted versus 
$m_\p L$
(solid curve)
is the ratio of the finite volume effect
$\D \be_M$ 
to the infinite volume polarizability 
$\be_M$
given in 
Eq.~\eqref{eq:beta/beta}. 
Additionally shown (dashed curve) is the asymptotic volume formula
Eq.~\eqref{eq:betaasymp}.
} 
\label{fig:polarizability}
\end{figure}

In the regime where 
$QB / m_\p^2 \ll 1$, 
one can expand the finite volume effect 
$\D E$
order-by-order in 
$B^2$. 
This corresponds to a large transverse area
$A_\perp$
expansion, 
carried out at fixed longitudinal size 
$L_3$. 
The term at zeroth-order in 
$B^2$
is part of the finite volume correction to the neutral pion mass.%
\footnote{
Recall that volume averaging and evaluation at zero magnetic field are operations that do not commute. 
Above, we take the source average first, then evaluate at zero magnetic field. 
The finite volume effect is different than obtained by first restricting to zero magnetic field.
That restriction gives the $p$-regime formula.}
The second-order term gives a finite volume correction to the magnetic polarizability.  
Using the definition 
\begin{eqnarray}
E 
= 
m_\pi
- \frac{1}{2} \beta_M B^2 
+ 
\c O(B^4)
,\end{eqnarray}
we reproduce the infinite-volume polarizability of the neutral pion%
~\cite{Bijnens:1987dc,Donoghue:1988eea,Holstein:1990qy}
\begin{equation}
\be_M
= 
\frac{Q^2}{6 m_\pi (4 \pi F_\p)^2}
,\label{eq:betainf}
\end{equation}
by expanding the infinite-volume energy 
$E$
in Eq.~\eqref{eq:E}
to
$\c O(B^2)$. 
Carrying out the same expansion on 
$\D E$
in Eq.~\eqref{eq:DE}, 
we obtain the finite volume correction
\begin{eqnarray}
\D \be_M \big/ \be_M
=
2 \sum_{\n = 1}^\infty \n m_\pi  L_3 \, \, K_1 ( \n m_\pi  L_3)
,\label{eq:beta/beta}
\end{eqnarray}
where 
$K_1(z)$
is a modified Bessel function. 
As 
$\D \be_M > 0$,  
the finite volume leads to a greater magnetic susceptibility of the neutral pion. 
For asymptotically large volumes, 
we have the fractional volume effect
\begin{equation}
\D \be_M \big/ \be_M
=
\sqrt{2 \pi m_\pi L_3}  \, \, e^{ - m_\pi L_3}
+ 
\cdots
\label{eq:betaasymp}
.\end{equation}
Taking 
$L_3 = L$, 
the finite volume effect on the magnetic polarizability of the neutral pion is shown in 
Fig.~\ref{fig:polarizability}, 
and contrasted with its asymptotic behavior. 
One needs 
$m_\p L > 4$
to have a volume effect of 
$< 10\%$.

\section{Summary of Key Results}
\label{sec:conclusion}

Chiral perturbation theory gives a model-independent description of low-energy QCD. 
We utilize this effective theory to determine finite volume effects on QCD observables in a uniform magnetic field. 
Our attention is restricted to charge neutral observables at next-to-leading order in the chiral expansion, 
including:
the chiral condensate, magnetization, pressure anisotropy, and neutral pion effective action. 
These observables receive loop corrections from charged pions, 
which are subject to magnetic periodic boundary conditions. 
Due to magnetic periodicity, 
the finite-volume theory retains a remnant of the continuous translational invariance of the Landau level problem in infinite volume. 
This remnant translational invariance has a desirable feature for charged pion loop contributions: 
spacetime averaging produces the infinite transverse-area results at a given magnetic field, 
namely
$A_\perp \to \infty$ 
with 
$Q B$ 
held fixed. 
Finite volume effects of suitably averaged quantities thus depend on the transverse size only through the magnetic field quantization 
condition, 
not through transverse images. 
Such averaging, 
moreover, 
is essentially a standard part of nearly all lattice QCD calculations of these observables.

A summary of key results is as follows.

\begin{enumerate}[label=(\roman*),noitemsep]
\item 
Finite volume corrections to the chiral condensate in a magnetic field are computed in Sec.~\ref{s:cond}. 
The chiral condensate in finite volume is spatially varying due to the remnant translational invariance, 
as shown in 
Fig.~\ref{fig:ccosc}. 
The spatially averaged condensate, 
however, 
receives only image corrections from the longitudinal direction, 
as given by 
Eq.~\eqref{eq:DRbar}. 
The finite volume effect on the magnetic field dependence of the condensate is shown in 
Fig.~\ref{fig:ccavg}. 
For the smallest magnetic flux quanta, 
the effect can be 
$\gtrsim 10\%$, 
even for 
$m_\p L = 4$. 
The finite volume serves to further catalyze chiral symmetry breaking in a magnetic field.

\item 
We compute the magnetic pressure anisotropy using chiral perturbation theory in Sec.~\ref{s:aniso}.
The pressure anisotropy at fixed flux can be used on the lattice to determine the magnetization of the QCD vacuum%
~\cite{Bali:2013esa}. 
In finite volume, 
the matter contribution to the pressure anisotropy depends on the magnetization, 
but there is an additional term that is purely a finite volume artifact, 
see Eq.~\eqref{eq:dp}. 
The finite volume effect on the pressure anisotropy is shown in 
Fig.~\ref{fig:anisotropy}. 
The additional term leads to the dominant finite volume effect, 
which is quite substantial. 
The finite volume produces a larger pressure anisotropy, 
consequently a larger apparent magnetization (in magnitude).

\item 
The neutral pion in a magnetic field is taken up in Sec.~\ref{sec:action}.
Calculation of the neutral pion effective action in finite volume is efficaciously performed using coordinate-space methods
pioneered by Schwinger. 
The end result 
Eq.~\eqref{eq:EffA2}
features a coordinate-dependent potential for the neutral pion that complicates the behavior of its correlation function
Eq.~\eqref{eq:G00}. 
Even after projecting the sink onto vanishing momentum, 
the two-point function generally depends on the location of the source. 
Fortunately, 
this finite volume effect is estimated to be quite small, 
as shown in 
Fig.~\ref{fig:correlator}.

\item 
While the coordinate-dependent effective potential of the neutral pion leads to momentum non-conservation between the source and sink, 
source averaging is carried out in practice to mitigate gauge noise. 
The effect of source averaging is to approximately project the correlation function to zero momentum at the source, 
thus restoring translational invariance. 
With momentum conserved, 
the two-point function takes a simple form
Eq.~\eqref{eq:C00}, 
from which the finite volume correction to the neutral pion energy can be identified. 
One application of this result is the determination of the finite volume effect 
on the magnetic polarizability of the neutral pion, 
which is shown in 
Fig.~\ref{fig:polarizability}
and points to a 
$10 \%$
effect for 
$m_\p L = 4$. 

\end{enumerate}

\noindent
Finally, although our investigation concerns the finite-volume effects in lattice QCD calculations, 
our results hint at potentially important finite-size effects due to field inhomogeneities. 
It would be interesting to adapt the techniques presented here to the study of  
magnetic fields of finite spatial extent, which could be relevant for heavy ion collisions.

\acknowledgments

P.A. acknowledges the hospitality of The City College of New York, and the Graduate Center of The City University of New York. 
P.A. also acknowledges the support of the Kavli Institute for Theoretical Physics, Santa Barbara, 
through which the research was supported in part by the National Science Foundation under Grant No. 
NSF PHY-1748958.

\appendix

\section{Finite Volume Formulas}
\label{s:A}

For numerical evaluation, 
the coincident charged pion propagator in 
Eq.~\eqref{eq:Gbar+}
is best expressed in terms of 
Jacobi elliptic-theta functions. 
Due to the finite volume Aharonov-Bohm factor
$(-1)^{N_\Phi \n_1 \n_2}$, 
the propagator generally requires three of the four canonically numbered functions 
\begin{eqnarray}
\vartheta_2 (z,q) 
&=&
2 \sum_{\n = 0}^\infty 
\cos \left[ (2 \n + 1) z \right]
q^{(\n+\frac{1}{2})^2}
,\notag \\
\vartheta_3 (z,q) 
&=&
1 + 2 \sum_{\n = 1}^\infty 
\cos \left( 2 \n z \right)
q^{\n^2}
,\notag \\
\vartheta_4 (z,q) 
&=&
1 + 2 \sum_{\n = 1}^\infty 
(-1)^\n
\cos \left( 2 \n z \right)
q^{\n^2}
,\end{eqnarray}
and expressions depend on whether the flux quantum
$N_\Phi$
is even or odd. 
There is one elliptic-theta function required to express the image sum for each direction. 
The coincident propagator of the charged pion is written in terms of a modification factor
$\Theta (x_\perp | s)$
in 
Eq.~\eqref{eq:G+fin}. 
This factor contains the image sums in the form
\begin{eqnarray}
\Theta(x_\perp | s)
&=&
\Theta_\parallel(s) \, \Theta^{(N_\Phi \text{mod } 2)}_\perp (x_\perp | s)
\label{eq:Thetaperp}
.\end{eqnarray}
Image sums in the Euclidean time and magnetic-field directions contribute the factor 
\begin{equation}
\Theta_\parallel(s)
=
\vartheta_3 \big(0,e^{-\frac{\beta^2}{4s}} \big)
\vartheta_3 \big(0,e^{-\frac{L_3^2}{4s}} \big)
\label{eq:Thetaparallel}
,\end{equation}
whereas the sums over images transverse to the field direction produce oscillatory dependence on the transverse coordinates
through the function 
$\Theta^{(N_\Phi \text{mod } 2)}_\perp (x_\perp | s)$, 
which depends on whether the flux quantum
$N_\Phi$
is even or odd. 
In the former case, 
one has
\begin{eqnarray}
\Theta^{(0)}_\perp(x_\perp | s)
&=&
\vartheta_3 \big(\tfrac{Q \theta_1}{2} + \tfrac{\p N_\Phi x_2}{L_2}, e^{-\frac{Q B L_1^2}{4 \tanh QB s}} \big)
\notag \\
&& \phantom{S} \times 
\vartheta_3 \big(\tfrac{Q \theta_2}{2} - \tfrac{\p N_\Phi x_1}{L_1}, e^{-\frac{Q B L_2^2}{4 \tanh QB s}} \big)
,\quad
\end{eqnarray}
while in the latter case, 
one arrives at two different contributions
\begin{eqnarray}
\Theta^{(1)}_\perp(x_\perp | s)
&=&
\vartheta_3 \big(Q \theta_1 + \tfrac{2 \p N_\Phi x_2}{L_2}, e^{-\frac{Q B L_1^2}{\tanh QB s}} \big)
\notag \\
&& \phantom{S} \times
\vartheta_3 \big(\tfrac{Q \theta_2}{2} - \tfrac{\p N_\Phi x_1}{L_1}, e^{-\frac{Q B L_2^2}{4 \tanh QB s}} \big)
\notag \\
&& 
+ \,
\vartheta_2 \big(Q \theta_1 + \tfrac{2 \p N_\Phi x_2}{L_2}, e^{-\frac{Q B L_1^2}{\tanh QB s}} \big)
\notag \\
&& \phantom{S} \times
\vartheta_4 \big(\tfrac{Q \theta_2}{2} - \tfrac{\p N_\Phi x_1}{L_1}, e^{-\frac{Q B L_2^2}{4 \tanh QB s}} \big)
,\quad
\end{eqnarray}
after separating the even and odd images in the $\hat{x}_1$-direction. 
Identical results are obtained from an analogous expression that separates the even and odd images in the 
$\hat{x}_2$-direction. 
This is due to the readily proven identity
\begin{eqnarray}
&&
\vartheta_3 (2 z_1, q_1^4) \, \vartheta_3 (z_2, q_2)
+
\vartheta_2 (2 z_1, q^4_1) \, \vartheta_4 (z_2, q_2)
=
\notag \\
&&
\vartheta_3 (2 z_2, q_2^4) \, \vartheta_3 (z_1, q_1)
+
\vartheta_2 (2 z_2, q^4_2) \, \vartheta_4 (z_1, q_1)
.\end{eqnarray}

Results for the neutral pion propagator
can be obtained from those of the charged pion by evaluation at 
$Q = 0$. 
The zero-charge limit produces the coincident finite-volume propagator of the neutral pion
$\ol G_0(0,0)$ 
in 
Eq.~\eqref{eq:G0fin}.  
In that case, 
the image sums are contained in the function 
$\Theta_0(s)$, 
which is defined as
\begin{equation}
\Theta_0 ( s ) 
=
\prod_{\mu=0}^3 
\vartheta_3 \Big( 0, e^{- \frac{L_\m^2}{4s}} \Big)
\label{eq:Theta0s}
.\end{equation}
At zero temperature,
one has 
$\beta = \infty$
and consequently
$\vartheta_3(0,0) = 1$.
The product over all
$\m$ 
is then reduced to that over the three spatial directions, 
and reflects that only 
$\n_0 = 0$
contributions remain. 
This feature is shared by the charged pion propagator, 
due to the temperature dependence of 
Eq.~\eqref{eq:Thetaparallel}.

\section{Covariant Derivatives of the Coincident Propagator}
\label{s:B}

In computing the effective action of the neutral pion in 
Sec.~\ref{s:Comp}, 
an additional one-loop diagram element is required beyond the coincident propagators given in 
Sec.~\ref{s:IIC}.
The new element is the charged pion tadpole diagram with two covariant derivatives of the pion propagator, 
specifically of the form 
\begin{equation}
D^2 G_+ (x,x)
\equiv
\lim_{x' \to x}
\big\langle \, D'_\mu \pi^{+}(x')  D_\mu \pi^{-} (x) \, \big\rangle 
.\label{eq:DDG+} 
\end{equation}
Point splitting has been introduced to regulate an ultraviolet divergence in a way consistent with dimensional regularization. 
As this divergence is independent of both the magnetic field and the volume, 
it would nevertheless be removed in the zero-field, infinite-volume renormalization scheme that we employ.

While the covariant derivatives of the charged pion propagator are straightforward to evaluate, 
a more circuitous route to compute 
Eq.~\eqref{eq:DDG+}
proves beneficial. 
One notes that
$[D'_\m, D_\m ] =0$
for each 
$\m$, 
thus allowing us to arrive at the identity 
\begin{equation}
D'_\m D_\m =  - \tfrac{1}{2} \left( D'_\m D'_\m + D_\m D_\m \right) + \tfrac{1}{2} (D'_\m + D_\m)^2 
.\end{equation}
Making use of this identity and the Green's function relation given in 
Eq.~(\ref{eq:GreenEQ}), 
we find  
\begin{equation}
\label{eq:DDG}
D^2 G_+ (x,x)
=
- m^2 G_+(x,x)
+ 
\tfrac{1}{2} \partial^2_\m G_+(x,x)
.\end{equation}
To obtain this relation, 
note that point splitting requires one to evaluate at 
$x' \neq x$
before taking the spacetime points to be coincident
$x' \to x$.
The relation in Eq.~\eqref{eq:DDG} agrees with that obtained by direct covariant differentiation, 
however, 
a proper-time integration by parts is needed to uncover the form given above. 
It is also possible to arrive at the relation by considering how the diagram element enters as a perturbative 
correction to the neutral pion Green's function Eq.~\eqref{eq:dG0}.  
In that case, 
one can perform integration by parts over the intermediate spacetime coordinate. 
Such integration by parts is efficacious for other contributions to the effective action, 
as discussed in Sec.~\ref{s:Comp}.

Using 
Eq.~\eqref{eq:DDG}, 
the new one-loop diagram element can be expressed in terms of the coincident charged pion propagator 
Eq.~\eqref{eq:coin}
and derivatives thereof.
The latter contribution is given by
\begin{eqnarray}
\partial^2_\m G_+(x,x)
=
-
\sum_{\n_\mu}
(QB \n_\perp L_\perp)^2 
f(\n_\perp, x_\perp)
\,
g_+(\n_\m)
,\,\,\,
\label{eq:ddG+}
\end{eqnarray}
and is written in terms of the functions 
$f(\n_\perp, x_\perp)$
and
$g_+(\n_\m)$
that appear in Eqs.~\eqref{eq:fperp} and \eqref{eq:g+}, 
respectively. 
Alternatively, 
this contribution can be expressed in terms of a proper-time integral
\begin{eqnarray}
\partial^2_\m G_+(x,x)
=
\int_0^\infty ds \, \frac{e^{- s m_\p^2}}{(4 \p s)^2}
\frac{QB s}{\sinh QB s}
\partial^2_\m \Theta(x_\perp,s)
,\,\,\,
\end{eqnarray}
where 
$\Theta(x_\perp,s)$
is given in Eq.~\eqref{eq:Thetaperp}. 
It is straightforward to differentiate this function, 
with the result producing dependence on various
$\vartheta''_j (z,q)$, 
where primes denote differentiation with respect to the first argument of 
the elliptic-theta functions.  
The expression for 
$\partial^2_\m \Theta(x_\perp,s)$
is quite lengthy, 
and is omitted for brevity.

\section{Two-Point Function Modification}
\label{s:C}

After projection onto zero spatial momentum at the sink, 
the neutral pion two-point function in 
Eq.~\eqref{eq:twopt}
can be cast in a form very similar to Eq.~\eqref{eq:G00}
\begin{eqnarray}
\c C_0 (T,x_\perp) 
&=&
\frac{e^{- E T}}{2 E}
\Big[
1 + \c R(T,x_\perp,\theta_\perp)
\Big]
,\end{eqnarray}
where, 
after integration over the intermediate time,  
the modification function 
$\c R(T,x_\perp,\theta_\perp)$
is given by
\begin{eqnarray}
\c R(T,x_\perp,\theta_\perp)
&=&
\sum_{\vec{p}_\perp}
\int_0^{L_\perp} d^2 y_\perp
\,e^{i \vec{p}_\perp \cdot (\vec{y} - \vec{x})_\perp} 
\,\c V(y_\perp)  
\notag \\
&& \phantom{space} \times
\frac{\frac{E}{\, E_{\vec{p}_\perp}} e^{(E - E_{\vec{p}_\perp}) T} {-}1}{\vec{p} \, {}_\perp^2}
,\end{eqnarray}
with 
$E_{\vec{p}} = \sqrt{E^2 + \vec{p}\, {}^2}$.
Explicit coordinate dependence of the effective potential 
$\c V(y_\perp)$
leads to 
non-conservation of momentum between the source and sink. 
Thus, 
there are contributions from momentum modes with
$\vec{p}_\perp \neq 0$. 
This applies to the charged pion contribution to the effective potential. 
By contrast, 
the neutral pion contribution
maintains momentum conservation, 
and can readily be included in the finite volume effect on the energy
$\D E$ 
in 
Eq.~\eqref{eq:G00}.

Writing the charged pion contribution to the effective potential 
Eq.~\eqref{eq:Vyperp} 
in terms of its image contributions, 
the transverse position integral appearing in 
$\c R(T,x_\perp,\theta_\perp)$
can be performed, 
because it is simply a Fourier transform
\begin{eqnarray}
\int_0^{L_\perp}
d^2 y_\perp \,
e^{i \vec{p}_\perp \cdot \vec{y}_\perp} 
f(\n_\perp, y_\perp)
&=&
\delta_{n_1, N_\Phi \n_2} 
\,
\delta_{n_2, - N_\Phi \n_1}
\notag \\
&& \times
\, A_\perp \, 
f(\n_\perp, 0)
,\end{eqnarray}
where the phase function 
$f(\n_\perp,x_\perp)$
is given in Eq.~\eqref{eq:fperp}. 
The transverse momentum sums are now trivial to perform, 
leading to the charged pion contribution
\begin{eqnarray}
\c R(T,x_\perp,\theta_\perp)
&=&
\sum_{\vec{\n} - \vec{0}}
\tfrac{\, m_\pi^2 + (Q B \n_\perp L_\perp)^2}{F_\p^2}
f(\n_\perp,0)
\, g_+(\vec{\n})
\notag \\
&& \times \,
e^{ - i \vec{p}_\n \cdot \vec{x}_\perp} \,
\frac{\frac{E}{E_{\vec{p}_\n}} e^{(E - E_{\vec{p}_\n}) T} {-} 1}{\vec{p} \, {}_\n^2}
,\label{eq:Rmess}
\end{eqnarray}
where 
$g_+(\n_\m)$
is given in Eq.~\eqref{eq:g+}, 
and the image-dependent momentum modes are specified by
$\vec{p}_\nu
= 
2 \pi N_\Phi
\left( \frac{\n_2}{L_1}, - \frac{\n_1}{L_2} \right)
$.

The charged pion contribution to the modification function can be additively decomposed into two terms
\begin{eqnarray}
\c R(T,x_\perp,\theta_\perp) 
&=& 
\c R_\parallel(T) 
+ 
\c R_\perp(T,x_\perp,\theta_\perp)
\label{eq:Rdecomp}
,\end{eqnarray}
where 
$\c R_\perp(T,x_\perp,\theta_\perp)$
is the contribution from all images with 
$\n_\perp \neq 0$, 
while 
$\c R_\parallel(T)$
is the contribution from all images with 
$\n_\perp = 0$. 
The former contribution does not exhibit any simplifications. 
It is given by a formula almost identical to 
Eq.~\eqref{eq:Rmess}, 
but with 
$\sum_{\vec{\n} - \vec{0}}$
replaced by
$\sum_{\vec{\n}, \n_\perp \neq 0}$, 
and is the 
$\c R_\perp(T,x_\perp,\theta_\perp)$
function appearing in 
Eq.~\eqref{eq:G00}.
For the twist angle and source location averaging discussed in 
Sec.~\ref{s:Two}, 
it is crucial to note that the dependence on twist angles enters via the phase function 
$f(\n_\perp,0) = (-1)^{N_\Phi \n_1 \n_2} \, e^{i Q \theta_\perp \cdot \n_\perp}$, 
whereas the transverse location of the source appears exclusively in the Fourier phase
$e^{ - i \vec{p}_\n \cdot \vec{x}_\perp}$. 
Both averages thus project onto the sector with
$\n_\perp = 0$;
consequently, 
these averages of 
$\c R_\perp(T,x_\perp,\theta_\perp)$
are zero.

By contrast, 
the charged pion contribution to
$\c R_\parallel(T)$
has 
$\vec{p}_\nu = 0$, 
for which momentum is conserved between the source and sink. 
In taking the limit of vanishing momentum, 
we obtain
\begin{eqnarray}
\c R_\parallel(T)
&=&
-
\left( T + \frac{1}{E} \right)
\frac{m_\p}{2 F_\p^2}
\sum_{\n_3 \neq 0}
g_+(\n_3)
,\end{eqnarray}
where we have approximated a multiplicative factor of 
$E$ 
by 
$m_\p$,
because the difference is 
$\c O(F_\p^{-4})$. 
The form of 
$\c R_\parallel(T)$ 
is exactly what one expects for a small correction 
$\D E$
to the energy 
$E$. 
In terms of a two-point function with energy 
$E + \D E$, 
we have 
\begin{equation}
\frac{e^{ - ( E + \D E) T}}{2 ( E + \D E)}
=
\frac{e^{ - E T}}{2 E}
\left[ 1 - \left( T + \frac{1}{E} \right) \D E \right] 
+ \cdots
.\label{eq:deltaE}
\end{equation}
Provided 
$\D E \, T \ll 1$
and
$\D E / E \ll 1$, 
one can treat
$\D E$
as a perturbative correction to the energy obtained from the time dependence of a two-point function with a simple exponential falloff. 
Hence, 
we identify 
$\D E$
of Eq.~\eqref{eq:G00}
with
\begin{eqnarray}
\D E 
=
\frac{m_\p}{2 F_\p^2}
\left[
\sum_{\n_3 \neq 0}
g_+(\n_3)
-
\frac{1}{2} \,
\ol G_0(0,0)
\right]
,\end{eqnarray}
which additionally includes the neutral pion contribution. 
This finite volume effect
is expressed as a proper-time integral in 
Eq.~\eqref{eq:DE}.

\bibliography{bib}

\end{document}